\documentclass[a4paper,11pt]{article}
\usepackage{jheppub} % for details on the use of the package, please
\usepackage{epsfig,multicol} 
\usepackage{amsmath}
\usepackage{amssymb}
\usepackage{subfigure}
\usepackage{here}
\usepackage{wrapfig}
\usepackage{graphicx}
\usepackage[T1]{fontenc} % if needed

\title{\boldmath Phase Transitions of a (Super) Quantum Mechanical Matrix Model with a Chemical Potential}

\author[a]{Takehiro Azuma,}
\author[b]{Pallab Basu}
\author[b]{and Prasant Samantray}

\affiliation[a]{Institute for Fundamental Sciences, Setsunan University, 
17-8 Ikeda Nakamachi, Neyagawa, Osaka, 572-8508, Japan}
\affiliation[b]{International Center for Theoretical Sciences, Tata Institute of Fundamental Research, Bangalore, 560089, India}
\emailAdd{azuma@mpg.setsunan.ac.jp}
\emailAdd{pallab.basu@icts.res.in}
\emailAdd{prshkumar@gmail.com}

\abstract{In this paper, we study the finite-temperature matrix quantum mechanics with chemical potential term linear in the single trace of U$(N)$ matrices, via Monte Carlo simulation. In the bosonic case, we exhibit the existence of the Gross-Witten-Wadia (GWW) type third-order phase transition. We also extend our studies to the model with the fermionic degrees of freedom employing the non-lattice simulation via Fourier expansion, and explore the possibilities that there is a phase transition between the gapped and ungapped phase both in the absence and presence of the chemical potential term. We make a comparison of the phase diagram between the bosonic and fermionic cases.
}

\begin{document} 
\maketitle
\flushbottom

\section{Introduction}
\label{sec:intro}
Unitary matrix models are ubiquitous in present day theoretical studies owing to its simplicity and analytical tractability in the context of modeling complex physical systems. The problem of studying phase transitions in a theory of quantum gravity is one such setting where matrix models have been extensively employed. This follows from the fact that in the large-$N$ limit, matrix models are actually string theories in disguise. In fact matrix models at present are the most potential candidates for M-Theory which is a non-perturbative formulation of string theory. 

In particular, the $0+1$ dimensional matrix model can be thought of as the dimensionally reduced version of the ten-dimensional SYM theory. At finite temperature, the $0+1$ dimensional quantum theory can often be  characterized by a SU$(N)$ matrix, which in turn is the Polyakov lines. It has been conjectured that a certain supergravity solution in the decoupling limit is dual to a $0+1$ super quantum mechanical matrix model with sixteen supercharges \cite{Itzhaki:1998dd, 0707_4454}. Additionally, authors in ref. \cite{hep-th0406210} have investigated the thermal behavior of $1+1$ dimensional SU$(N)$ gauge theory on a circle of circumference $L$ in the context of Gregory-Laflamme transition near the horizon of D$0$-branes on a circle. They conclude that for the temperature $T$ in the regime $L^3 \lambda T <<1 ;~T^3 L >> \lambda$, where $\lambda$ is the usual 't Hooft coupling, the eigenvalues tend to concentrate around a point on the unit circle. Increasing the temperature further would make the eigenvalues spread out by filling up the entire circle via a GWW type black hole $\rightarrow$ string phase transition. However, the authors point out that at higher temperatures perturbation theory becomes suspect and the theory effectively becomes a $0+1$ dimensional matrix theory compactified over a circle of circumference $L$ - thus exhibiting a GWW type phase transition near $L^3 \lambda T \sim \textrm{O}(1)$.

GWW transition in the context of a gauge theory was considered in ref. \cite{0710_5873,AlvarezGaume:2006jg,hep-th0502227} to understand the string $\rightarrow$ black hole transition. By using the AdS/CFT correspondence, this transition in a AdS$_5 \times$ S$^5$ spacetime was mapped to a GWW-type third-order transition in the boundary field theory living on S$^3 \times$ R. Because of the compactness of S$^3$, the boundary theory effectively reduces to a multi-trace unitary matrix model corresponding to the zero mode of Polyakov line - which in turn exhibits the GWW phase transition. Such a reduction to unitary matrix model is non-trivial especially in the strong coupling regime due to the Gregory-Laflamme transition for small black holes. However, even in this regime it has been conjectured that unitary matrix models are good effective descriptions. In fact, near the transition temperature in a double scaled region one can compute the o$(1)$ part of the effective boundary theory action in terms of universal function $F(t)$ characterized by the equation $\partial^2_{t} F(t) = - f^2 (t)$, where $f(t)$ is the Painleve II function and $t$ is a variable which scales as the factor $(T-T_c)N^{2/3}$. Such a universality arises since near the transition the ``critical" system is fully characterized by the power ``$2/3$" in the regime $(T-T_c) \sim N^{-2/3}$. Subsequently, the GWW phase transition is exhibited in such unitary matrix models as the system displays a cross over from a gapped to an ungapped phase of the eigenvalue distribution. Such results have been verified both analytically (in the case of simple single trace unitary matrix models) and numerically in other complicated situations. 

It is now a well-established result that single trace unitary matrix models undergo a third-order phase transition in the large-$N$ limit \cite{GWW1,GWW2,GWW3}. Such studies typically restrict themselves to the bosonic sector of the theory.  Taking this cue, in our work we consider a matrix model with fermionic degrees of freedom and look for signatures of phase transitions. Unfortunately, adding fermionic degrees of freedom renders the theory difficult to track from a analytical standpoint and therefore we numerically study the system. 

Additionally, unlike the previous studies, the temperature is not the only tunable parameter in such matrix theories and we can introduce a chemical potential term in the action. Introduction of a chemical potential breaks supersymmetry. For simplicity we consider a chemical potential linear in the single trace of U$(N)$ matrices. This is so since in the large-$N$ limit one can consider U$(N)$ as the gauge group instead of SU$(N)$, without having any Nambu-Goldstone mode. This like choosing chemical potential for a Polyakov loop and gravity dual of such a configuration is named a Hedgehog black hole \cite{Headrick}. One can also consider more complicated chemical potentials but as it would become evident, our choice suffices to extract the relevant physical behavior of the system without any loss of generality.  Matrix models are characterized by their distributions of eigenvalues, and our focus would be to look for gapped distributions of eigenvalues in a model with fermions which would correspond to the development of a phase transition. In the purely bosonic case, such a phase transition has been directly correlated to a black hole $\rightarrow$ string transition.

The introduction of a chemical potential term can also be understood from a conceptual standpoint as follows. In a typical field theory, the dynamics of the theory is usually governed by the saddle point of the action. However, in cases where the potential term exhibits exotic behavior (like a discontinuity in its slope etc.) farther away from the saddle point, the usual techniques of perturbation theory in the saddle point approximation fail to capture such phenomena. In this sense, introduction of a chemical potential term is akin to adding a ``source" term to the action wherein one can traverse the entire potential by modulating the source function. In other words, one can trace the thermal history of a finite temperature matrix model by having access to the entire phase space via the chemical potential.

The organization of our paper is as follows. In section \ref{sec_bosonic_model}, we first consider the purely bosonic matrix model in the presence of a chemical potential. The essential idea is to study the vacuum expectation value of the order parameter (which in our case is the path ordered Polyakov loop) and its variations with respect to the temperature and chemical potential. This would reveal the nature of phase transitions in such a model and we plot the same using Monte Carlo techniques. In section \ref{sec_fermionic_model}, we then extend this work by including fermionic degrees of freedom and find that there is a phase transition between the gapped and ungapped phase both in the absence and presence of the chemical potential term. 
%We compare the nature of phase transitions in the fermionic model with that of the purely bosonic model.
In section \ref{sec_phase_diag} we discuss the phase diagram of the bosonic and fermionic model, and section \ref{sec_conclusion} is devoted to conclusion and outlook.
% and conclude by commenting on the physics of the gravity theory dual to our models.

To start with, let us consider the purely bosonic matrix model in the presence of a chemical potential.  

\section{Phase Transitions in a Bosonic Matrix Model} \label{sec_bosonic_model}
\subsection{Bosonic Finite-temperature Matrix Model}
The action of the bosonic finite-temperature matrix model is
\begin{eqnarray}
 S_{\textrm{b}} &=& \frac{1}{g^2} \int^{\beta}_{0} \textrm{tr}  \left\{ \frac{1}{2} \sum_{\mu=1}^{D} (D_t X_{\mu} (t))^2 - \frac{1}{4} \sum_{\mu,\nu=1}^{D}  [X_{\mu} (t), X_{\nu} (t) ]^2 \right\} dt, \label{bosonic_action}
\end{eqnarray}
where $D_t$ is a covariant derivative $\displaystyle D_t X_{\mu}(t) = \partial_t X_{\mu} (t) - i [A(t), X_{\mu} (t)]$. $A(t)$ and $X_{\mu} (t)$ are $N \times N$ hermitian matrices. The indices $\mu,\nu = 1,2,\cdots,D$ are contracted by the Euclidean metric, and $D$ is the dimensionality of the model. We work in units $g^2 N=1$. The Euclidean time $t$ has a finite extent $\beta$, which corresponds to the inverse temperature $\displaystyle \beta = \frac{1}{T}$.
Especially for $D=9$, this bosonic model is the high-temperature limit of the $(1+1)$-dimensional maximal super-Yang-Mills theory. The bosonic model has been so far studied analytically and numerically, for example, in refs. \cite{hep-th0406210,hep-th0508077,0704_3183,0706_3517,0710_2188,0710_5873,0901_4073,0910_4526,1207_3323,1403_7764}. This model has a U$(N)$ gauge symmetry
\begin{eqnarray}
 X_{\mu} (t) \to g(t) X_{\mu} (t) g^{\dagger} (t), \ \ A(t) \to g(t) A(t) g^{\dagger} (t) + i g(t) \frac{d g^{\dagger} (t)}{dt}. \label{boson_gauge}
\end{eqnarray}
Also, the action (\ref{bosonic_action}) is invariant under the transformations
\begin{eqnarray}
 & & X_{\mu} (t) \to X_{\mu} (t) + x_{\mu} E, \label{x_trans} \\
 & & A(t) \to A(t) + \alpha (t) E. \label{a_trans}
\end{eqnarray}
$E$ is an $N \times N$ unit matrix. $x_{\mu}$ and $\alpha (t)$ are c-numbers, and there is no $t$-dependence in $x_{\mu}$. The fields obey periodic boundary condition 
\begin{eqnarray}
 A(t+\beta) = A(t), \ \ X_{\mu} (t + \beta) = X_{\mu} (t). \label{periodic_boundary}
\end{eqnarray}
We adopt the static diagonal gauge 
\begin{eqnarray}
 A (t) = \frac{1}{\beta} \textrm{diag} (\alpha_1, \alpha_2, \cdots, \alpha_N), \label{static_diagonal}
\end{eqnarray}
where $\alpha_k$ ($k=1,2,\cdots,N$) no longer depends on $t$ and has a periodicity $2\pi$. This yields the gauge-fixing term 
\begin{eqnarray} 
 S_{\textrm{g.f.}} = - \sum_{k,l=1, k \neq l}^{N} \log \left| \sin \frac{\alpha_k - \alpha_l}{2} \right|, \label{gauge_fixing}
\end{eqnarray}
which is derived in refs. \cite{hep-th0310286,hep-th0601170}. The model (\ref{bosonic_action}) has a confinement/deconfinement (referred to as ``CD" henceforth) phase transition at a certain critical temperature.
The order parameter useful for studying the CD phase transition is 
\begin{eqnarray}
u_n =\frac{1}{N} \textrm{tr} U^n, \textrm{ where }  U = {\cal P} \exp \left( i \int^{\beta}_{0} A(t) dt \right). \label{u_n}
\end{eqnarray}
${\cal P}$ denotes the path-ordered product. In the static diagonal gauge (\ref{static_diagonal}) this is written as
\begin{eqnarray}
u_n = \frac{1}{N} \sum_{k=1}^{N} e^{in \alpha_k}. \label{u_n2}
\end{eqnarray}
 In ref. \cite{0910_4526}, using the large-$D$ expansion, they predicted at large $D$ that at $T=T_{c1}$ there is a second-order phase transition, and that at $T=T_{c2} > T_{c1}$ there is yet another phase transition of third order. In weakly coupled theory one may study various phase transitions analytically in a perturbative theory \cite{hep-th0310285, hep-th0508077}. It seems CD phase transition is of first order. There is no third-order GWW phase  transition. Only an unstable saddle point goes through a third-order GWW type phase transition. As we would discuss in the next chapter addition of a chemical potential changes this picture.
 
  In ref. \cite{1403_7764}, Monte Carlo simulation shows that at sufficiently small $D$ ($D \leqq 20$) the phase transition is of first order. Numerically, it is difficult to finally determine whether the phase transition is of first order or first$+$third order. Previous Monte Carlo studies have shown \cite{0706_3517,0901_4073,1207_3323,1403_7764} that some transitions, whose detail is not the issue of this paper, occur around the critical temperature $T_{c0}$ for the action $S_{\textrm{b}}$. This has been obtained as
\begin{eqnarray}
 \hspace*{-8mm} T_{c0} \simeq 1.32 \ (D=2), \ \ T_{c0} \simeq 1.10 \ (D=3), \ \ T_{c0} \simeq 0.95 \  (D=6), \ \ T_{c0} \simeq 0.90 \ (D=9). \label{critical0}
\end{eqnarray}

\subsection{Adding Chemical Potential to the Theory}

In the following, we study the saddle point of the gauge field by adding the chemical potential\footnote{In ref. \cite{0710_5873}, the overall coefficient was not $N \mu$ but $N
 \mu \beta$, and hence there is a difference in the notation.}
\begin{eqnarray}
 S_{\textrm{g}} = N \mu (\textrm{tr} U + \textrm{tr} U^{\dagger}) \label{chemical_pot}
\end{eqnarray}
Namely, we work on the action in the static diagonal gauge (\ref{static_diagonal}) under the unit $g^2 N=1$. In the weakly coupled $0+1$ bosonic theory (and also possibly in various similar models with a mass gap), there is a stable saddle point at low temperature and non-zero chemical potential. For suitable values of the parameters this saddle point may go through a GWW transition \cite{Basu:2005pj,Basu:2008uc}.  

Our action is
\begin{eqnarray}
 S &=& S_{\textrm{b}} + S_{\textrm{g}} + S_{\textrm{g.f.}} = N \int^{\beta}_{0} \textrm{tr} \left\{ \frac{1}{2} \sum_{\mu=1}^{D}   \left( D_t X_{\mu} (t) \right)^2 - \frac{1}{4} \sum_{\mu,\nu=1}^{D}  [X_{\mu} (t), X_{\nu} (t) ]^2 \right\} dt \nonumber \\
 & & \ \ + 2 N \mu \sum_{k=1}^{N} \cos \alpha_k - \sum_{k,l=1, k \neq l}^{N} \log \left| \sin \frac{\alpha_k - \alpha_l}{2} \right|. \label{our_bosonic_action}
\end{eqnarray}
The properties of the action $S_{\textrm{g}}$ without $S_{\textrm{b}}$ are presented in Appendix \ref{GWW_appendix}. The addition of $\displaystyle S_{\textrm{g}}$ breaks the invariance under (\ref{a_trans}) while the invariance under (\ref{x_trans}) is maintained. To put this action on a computer, we discretize the Euclidean time direction as
\begin{eqnarray}
 S_{\textrm{lat}} &=& N (\Delta t) \sum_{n=1}^{n_t}  \textrm{tr} \left\{ \frac{1}{2 (\Delta t)^2}  \sum_{\mu=1}^{D} (X_{\mu,n+1} - V X_{\mu,n} V^{\dagger})^2 - \frac{1}{4} \sum_{\mu,\nu=1}^{D} [X_{\mu,n}, X_{\nu,n} ]^2 \right\} \nonumber \\
 & & \ \ +  2 N \mu \sum_{k=1}^{N} \cos \alpha_k - \sum_{k,l=1, k \neq l}^{N} \log \left| \sin \frac{\alpha_k - \alpha_l}{2} \right|. \label{our_bosonic_action_lat}
\end{eqnarray}
$n_t$ is the number of lattice sites, and $\displaystyle (\Delta t) =\frac{\beta}{n_t}$ is the lattice spacing. \\ 
\noindent $\displaystyle X_{\mu,n} = X_{\mu} (t = n (\Delta t))$, and the boundary condition (\ref{periodic_boundary}) gives  $\displaystyle X_{\mu, n_t + 1} = X_{\mu,1}$.\\ 
\noindent $\displaystyle V = e^{i A (\Delta t)} = \textrm{diag} (e^{i \alpha_1/n_t}, e^{i \alpha_2/n_t}, \cdots, e^{i \alpha_n /n_t})$. The lattice size dependence turns out to be small enough, and we take $n_t =15$. We apply the Hybrid Monte Carlo (HMC) algorithm \cite{HMC_ref}\footnote{Yet another way to simulate the action (\ref{our_bosonic_action_lat}) is the heatbath algorithm, which is presented in Appendix B of ref. \cite{0706_3517}. Also, in ref. \cite{0901_4073}, instead of the lattice regularization (\ref{our_bosonic_action_lat}), they applied the Fourier expansion \cite{0706_1647,0707_4454} to the bosonic model, as well as the supersymmetric model.} to the action (\ref{our_bosonic_action_lat}).

\subsection{Results for the Bosonic Matrix Model}
\begin{figure}[htbp]
\centering % \begin{center}/\end{center} takes some additional vertical space
\includegraphics[width=6.6cm]{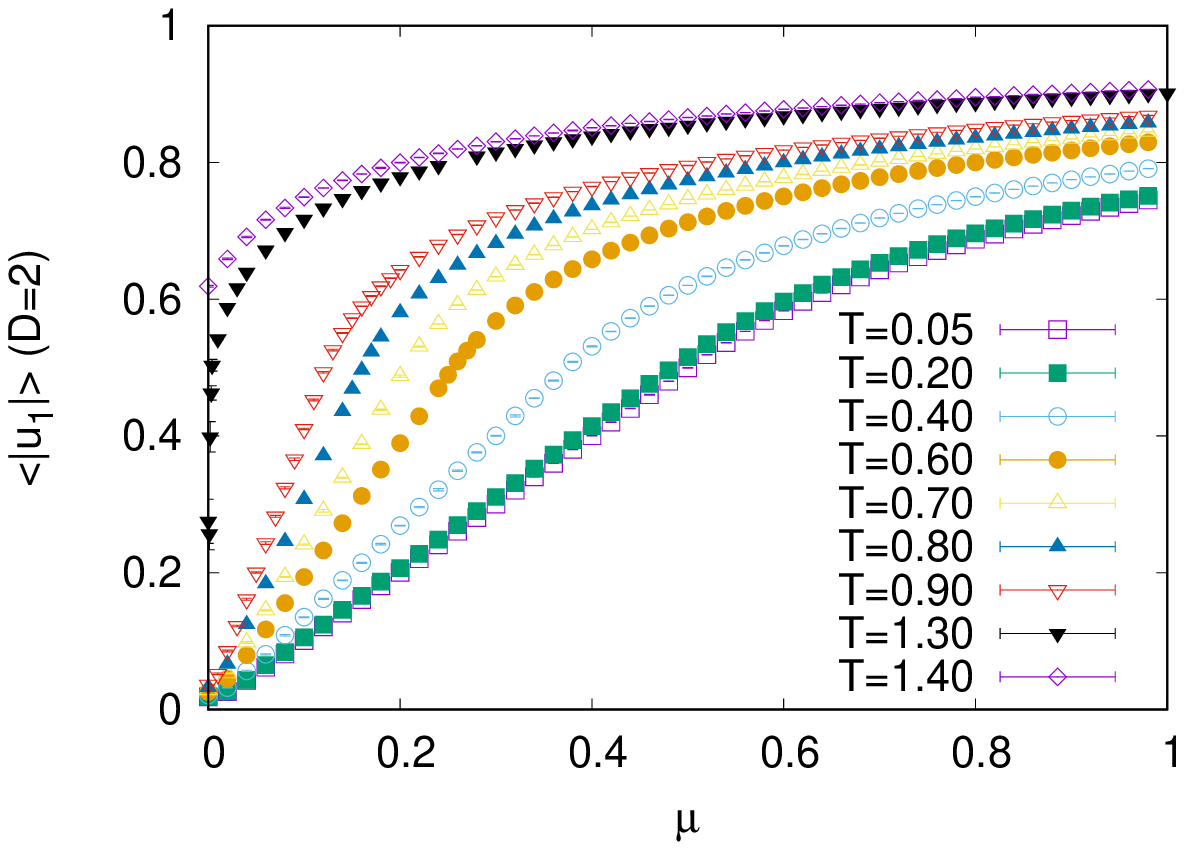}
\includegraphics[width=6.6cm]{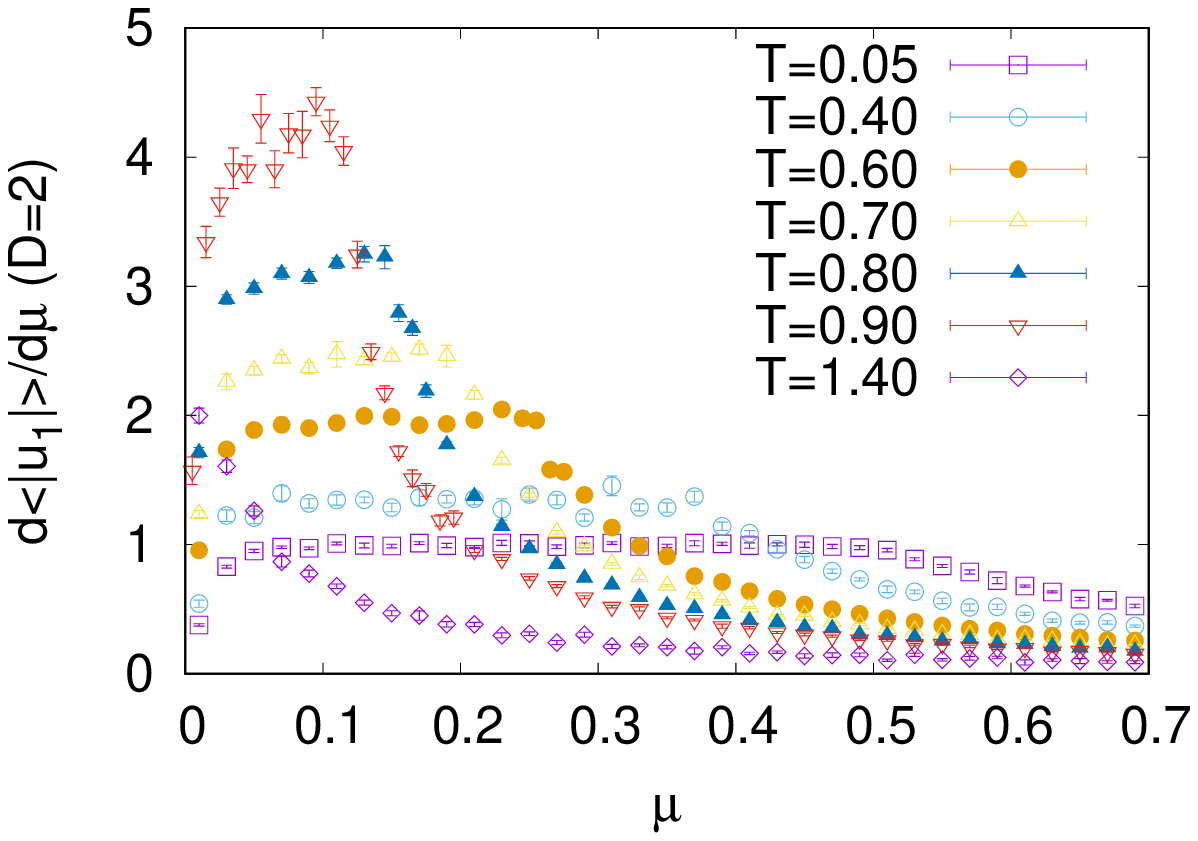}
\includegraphics[width=6.6cm]{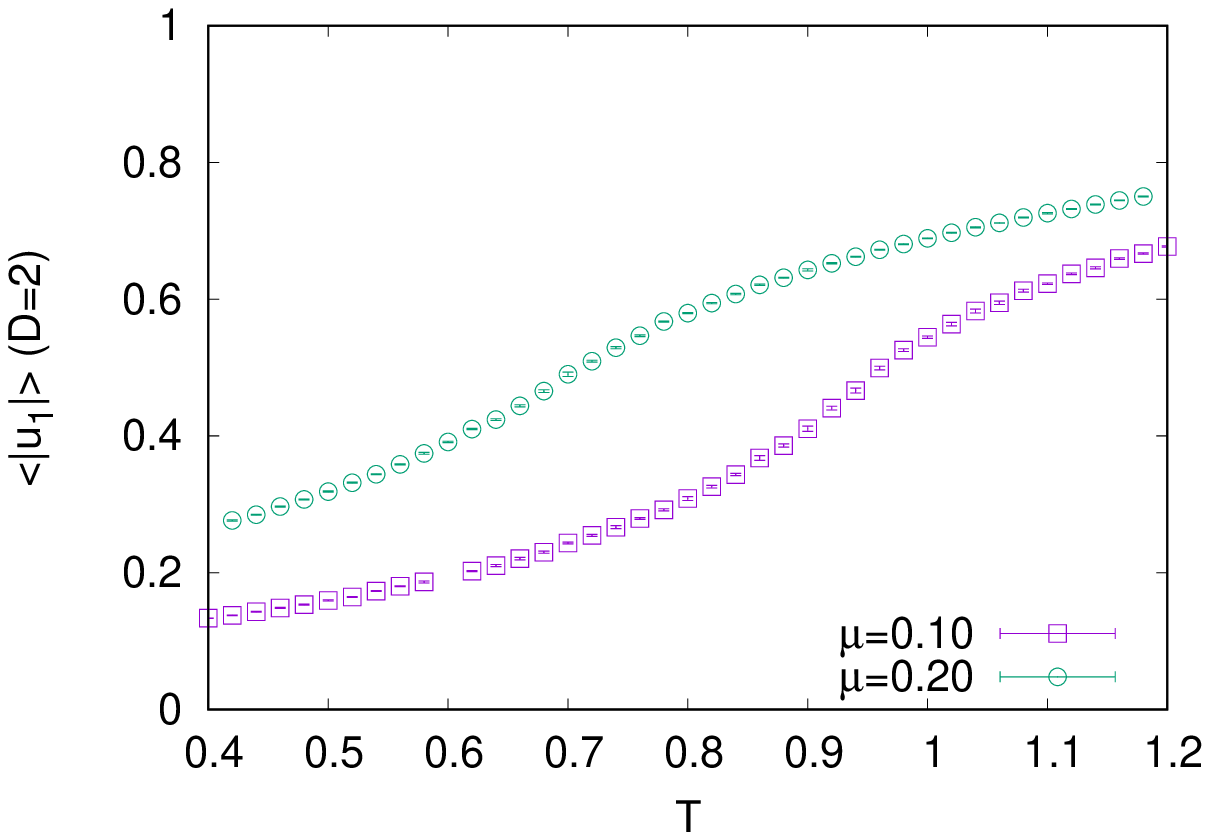}
\includegraphics[width=6.6cm]{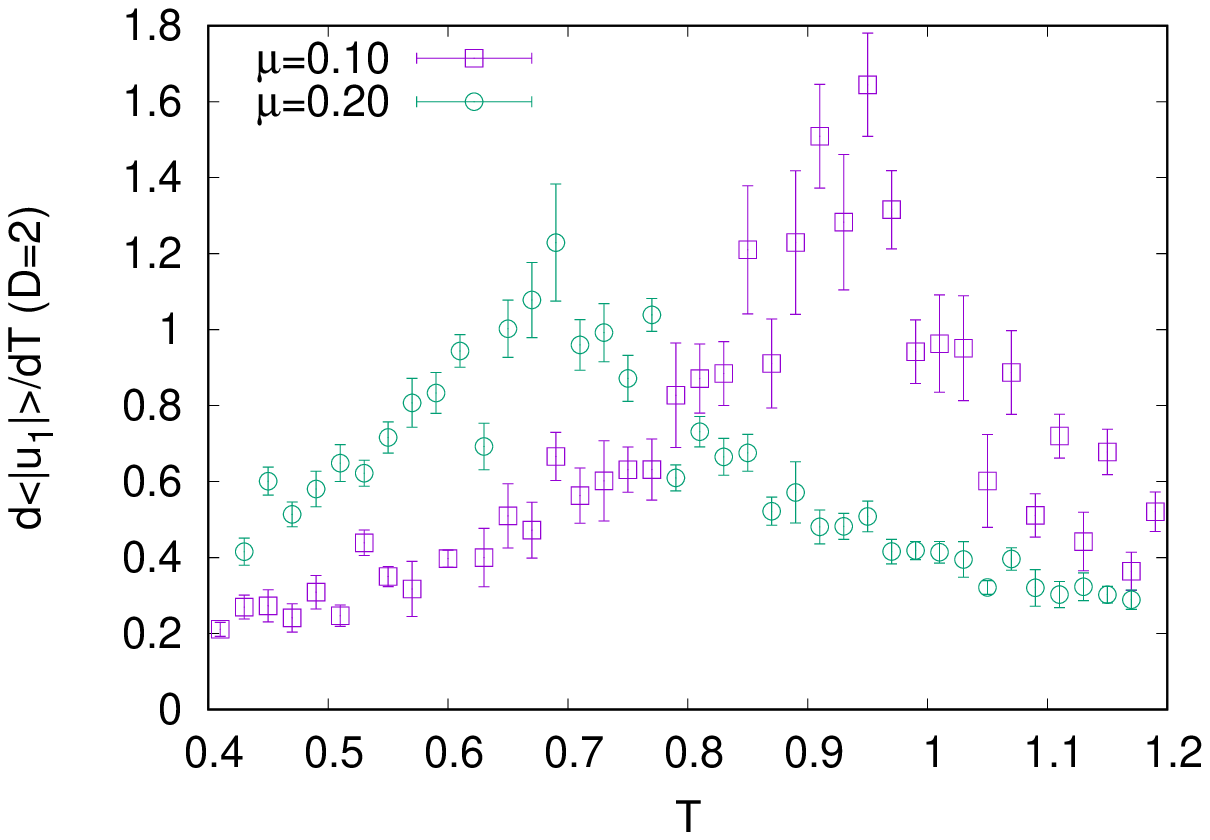}
%\vspace{4mm}
\caption{$\displaystyle \langle |u_{1}| \rangle$(left) and $\displaystyle \frac{d \langle |u_{1}| \rangle}{d \mu},\displaystyle \frac{d \langle |u_{1}| \rangle}{d T}$(right) for $D=2,N=48$ against $\mu$(top) and $T$(bottom).}
\label{un_data_d2}
\end{figure}

\begin{figure}[htbp]
\vspace*{-2mm}
\centering % \begin{center}/\end{center} takes some additional vertical space
\includegraphics[width=6.6cm]{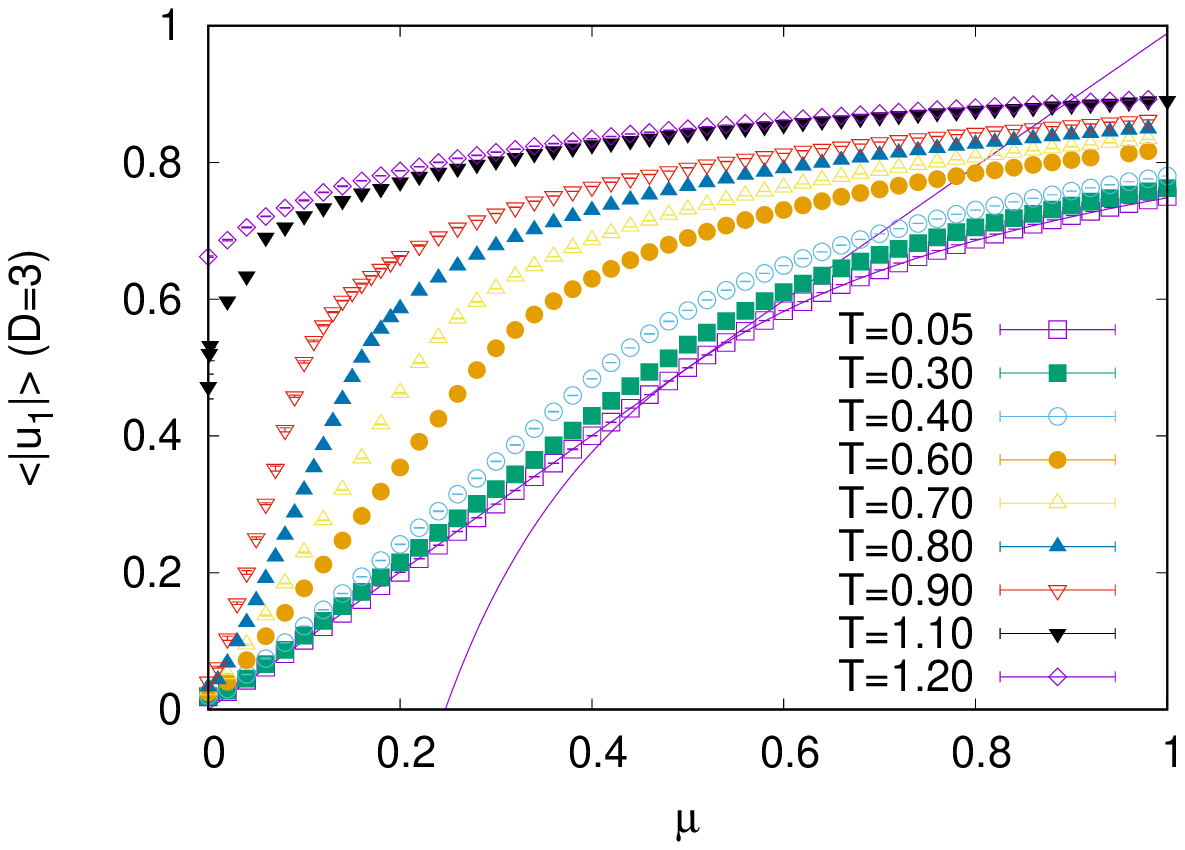}
\includegraphics[width=6.6cm]{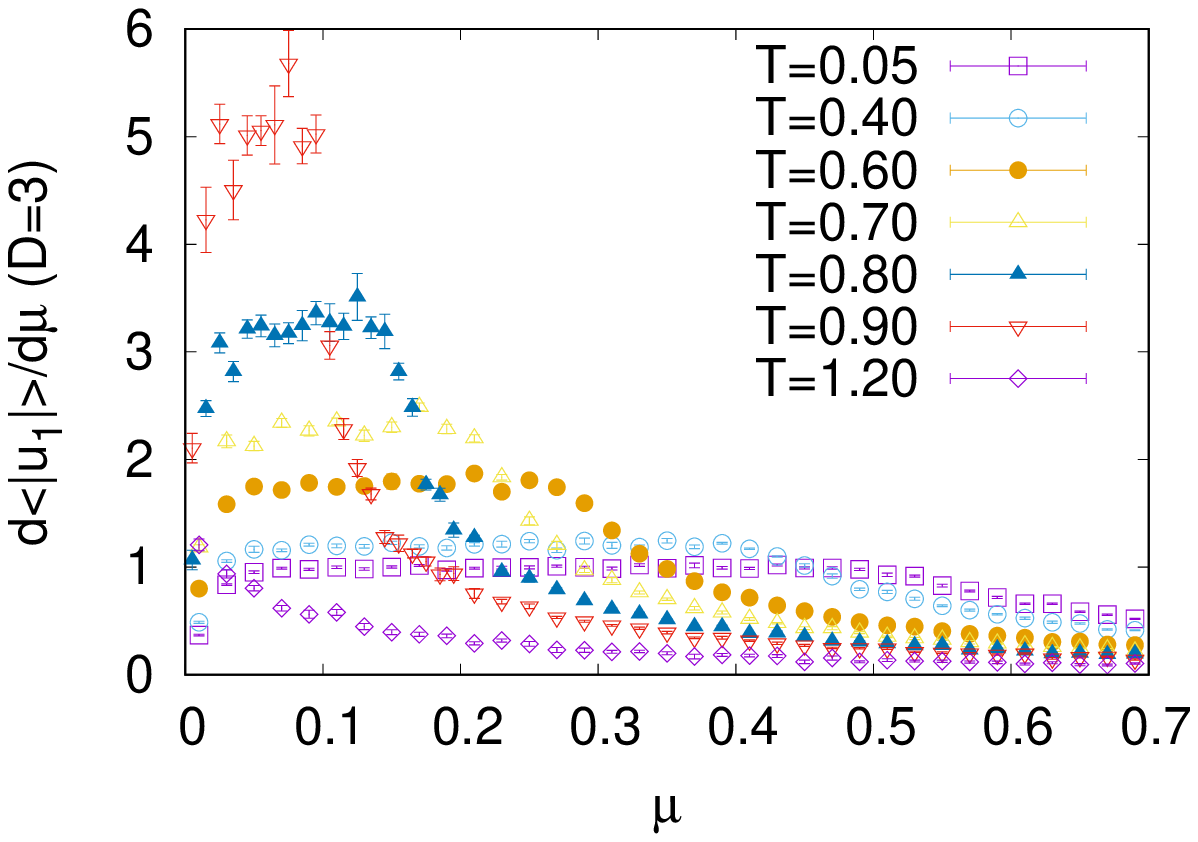}
\includegraphics[width=6.6cm]{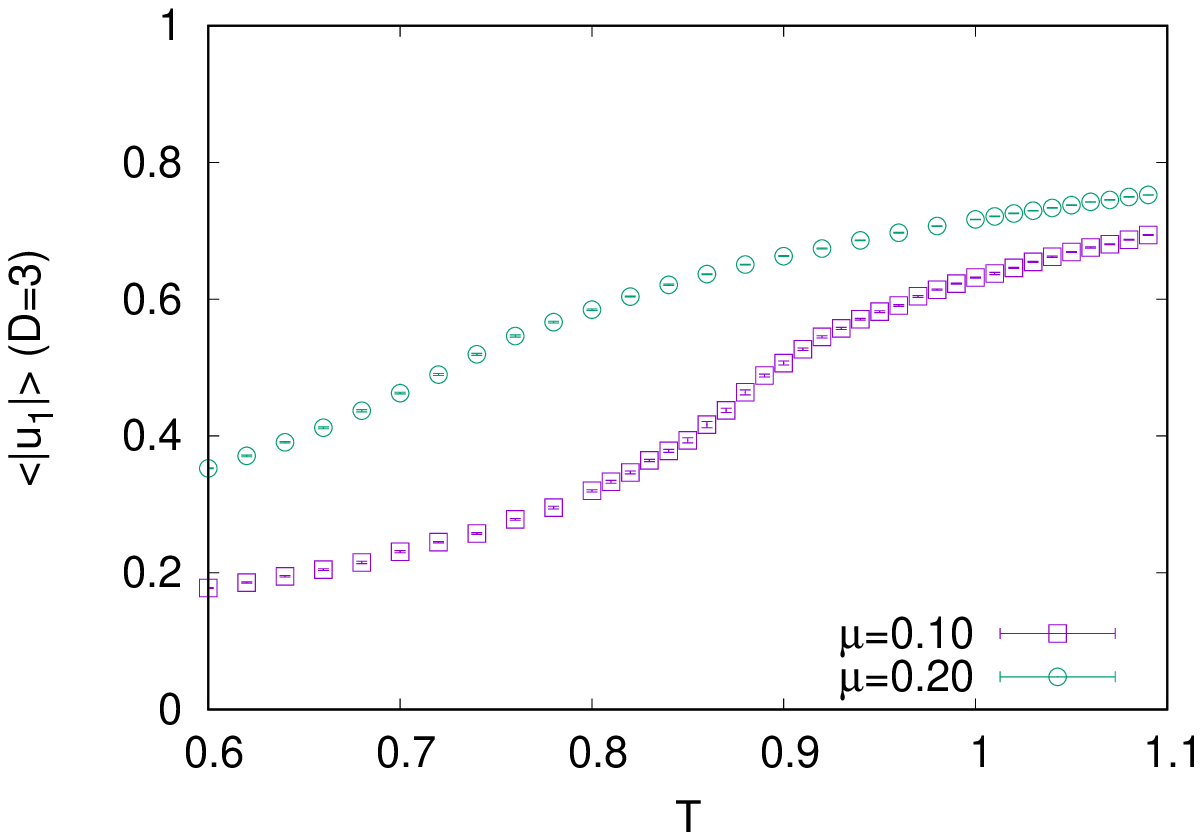}
\includegraphics[width=6.6cm]{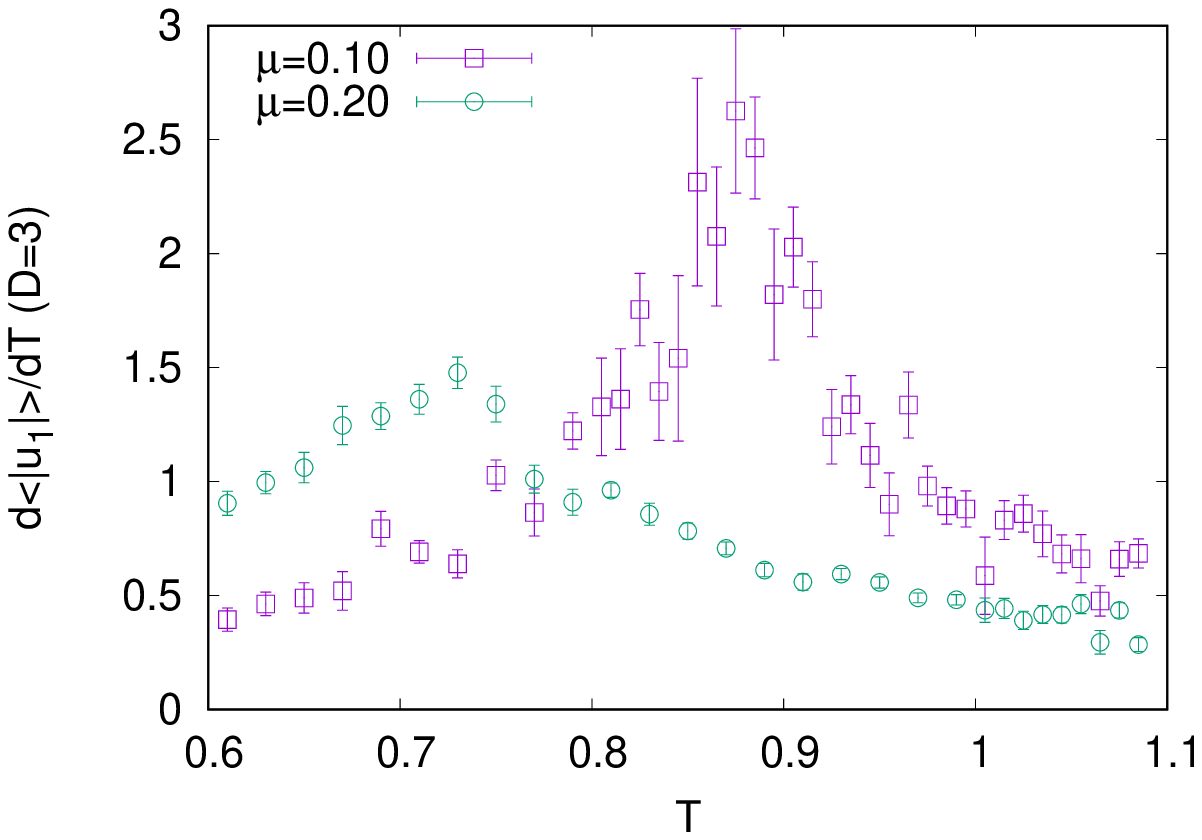}
%\vspace{4mm}
\caption{$\displaystyle \langle |u_{1}| \rangle$(left) and $\displaystyle \frac{d \langle |u_{1}| \rangle}{d \mu},\displaystyle \frac{d \langle |u_{1}| \rangle}{d T}$(right) for $D=3,N=48$ against $\mu$(top) and $T$(bottom).}
\label{un_data_d3}
\end{figure}

\begin{figure}[htbp]
\centering % \begin{center}/\end{center} takes some additional vertical space
\includegraphics[width=6.6cm]{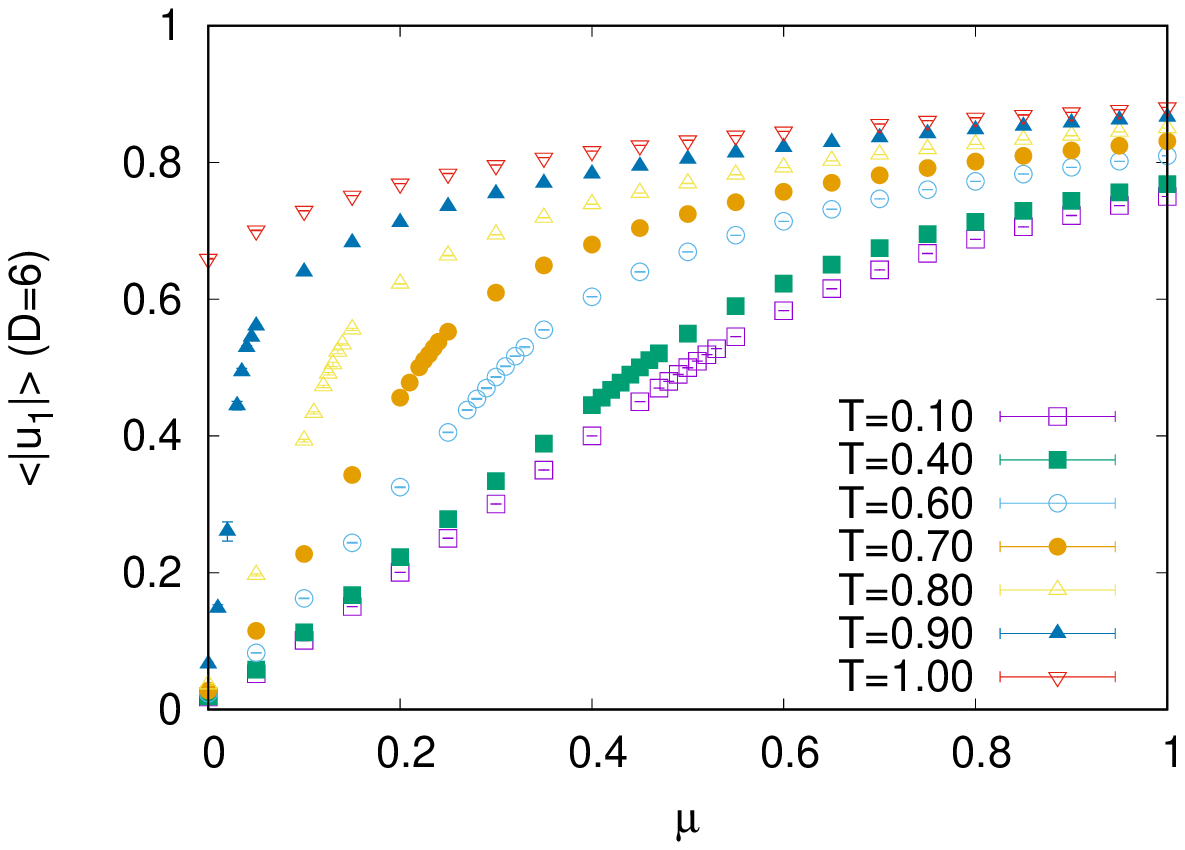}
\includegraphics[width=6.6cm]{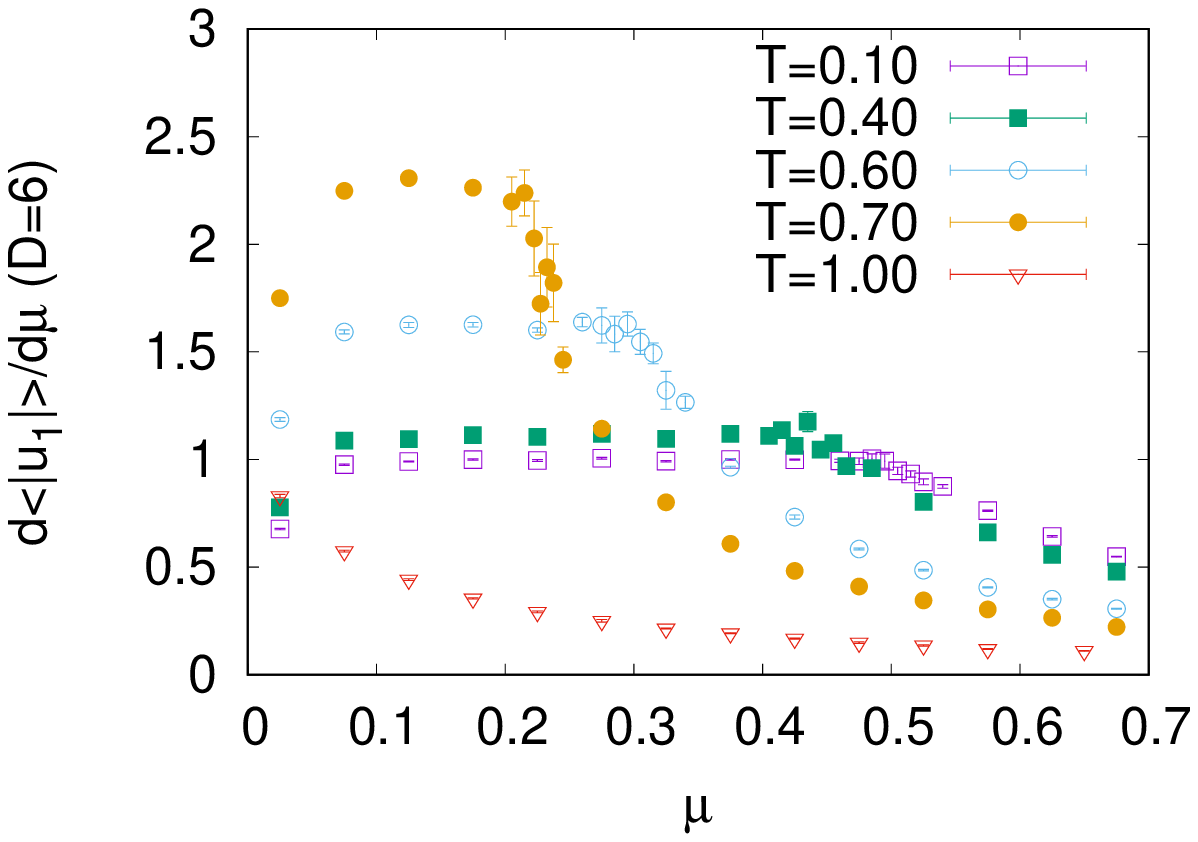}
\includegraphics[width=6.6cm]{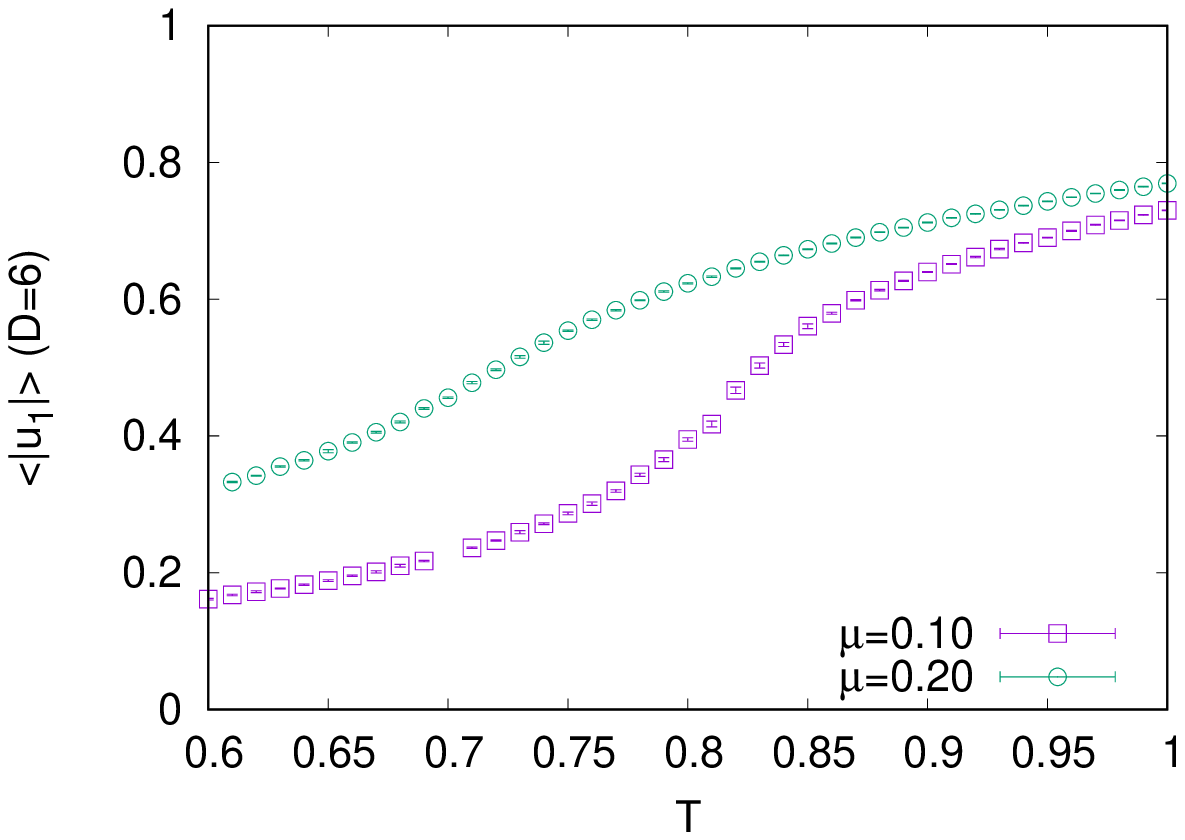}
\includegraphics[width=6.6cm]{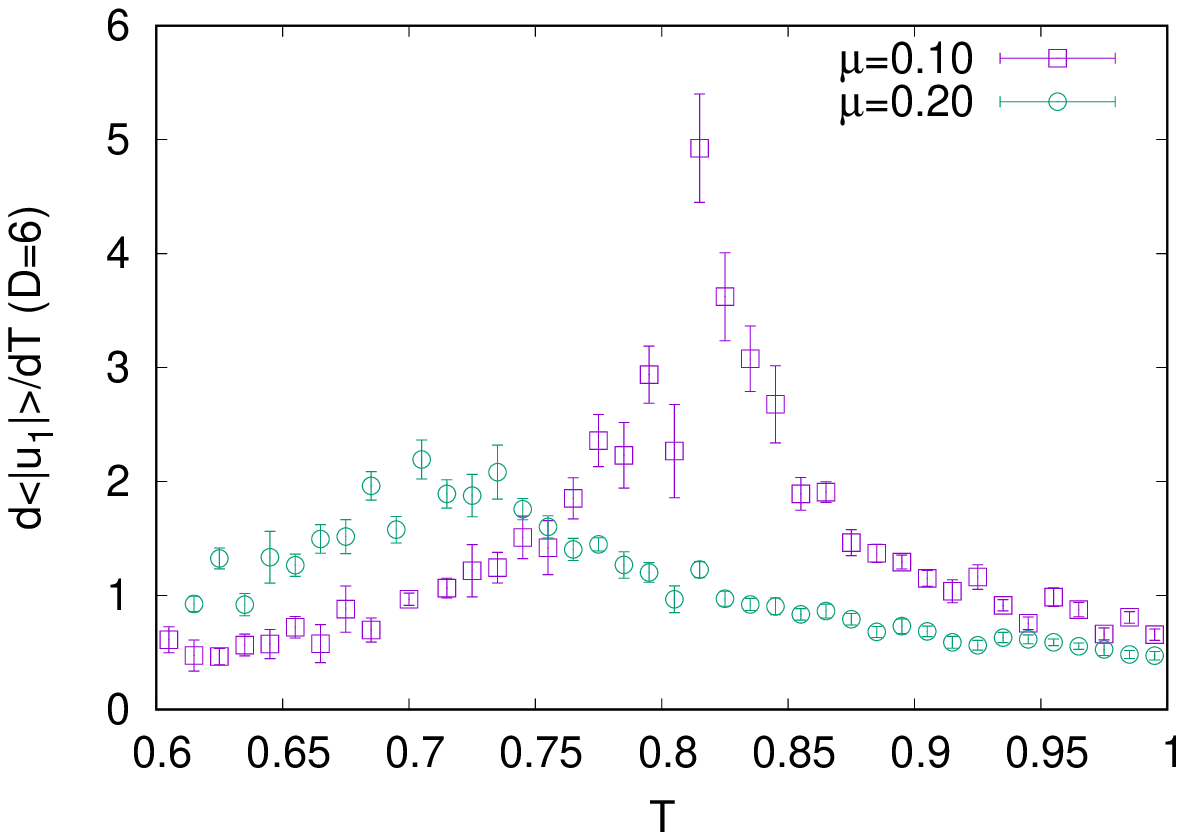}
%\vspace{4mm}
\caption{$\displaystyle \langle |u_{1}| \rangle$(left) and $\displaystyle \frac{d \langle |u_{1}| \rangle}{d \mu},\displaystyle \frac{d \langle |u_{1}| \rangle}{d T}$(right) for $D=6,N=48$ against $\mu$(top) and $T$(bottom).}
\label{un_data_d6}
\end{figure}

\begin{figure}[htbp]
\vspace*{-2mm}
\centering % \begin{center}/\end{center} takes some additional vertical space
\includegraphics[width=6.6cm]{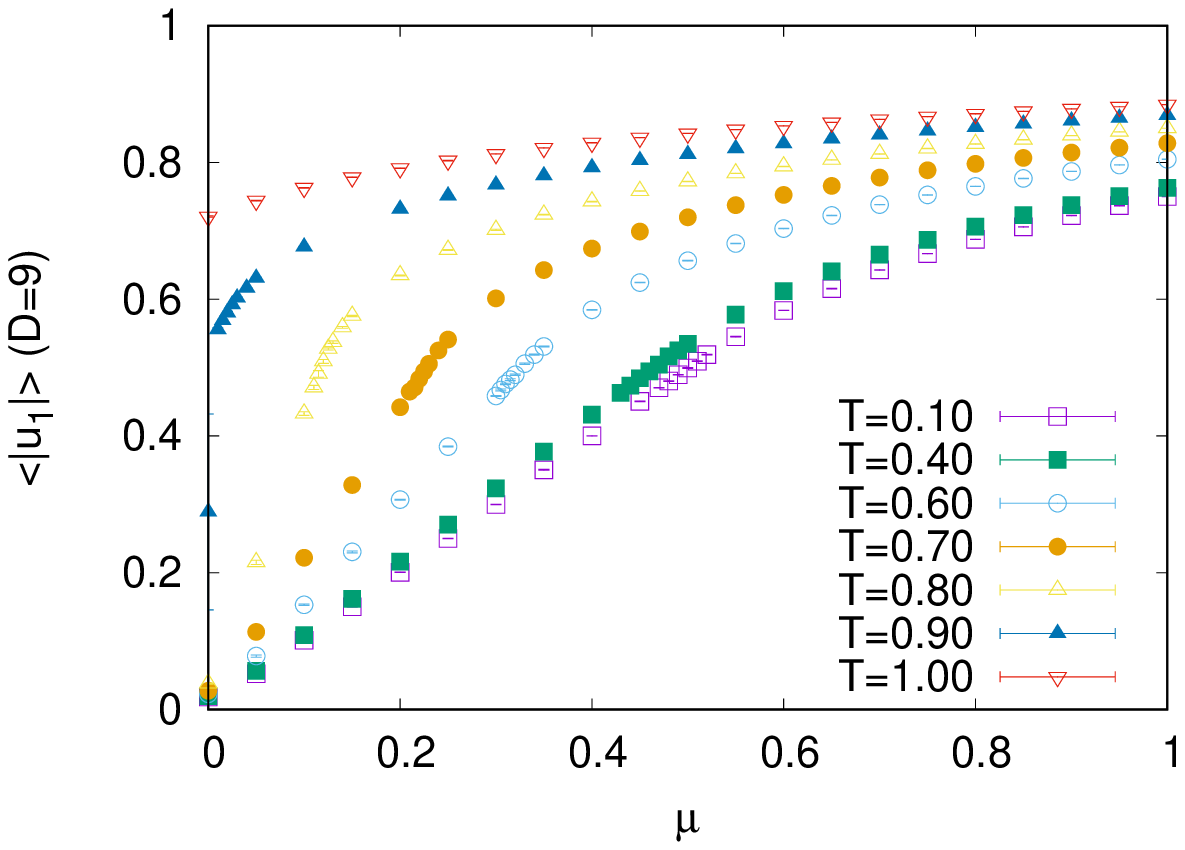}
\includegraphics[width=6.6cm]{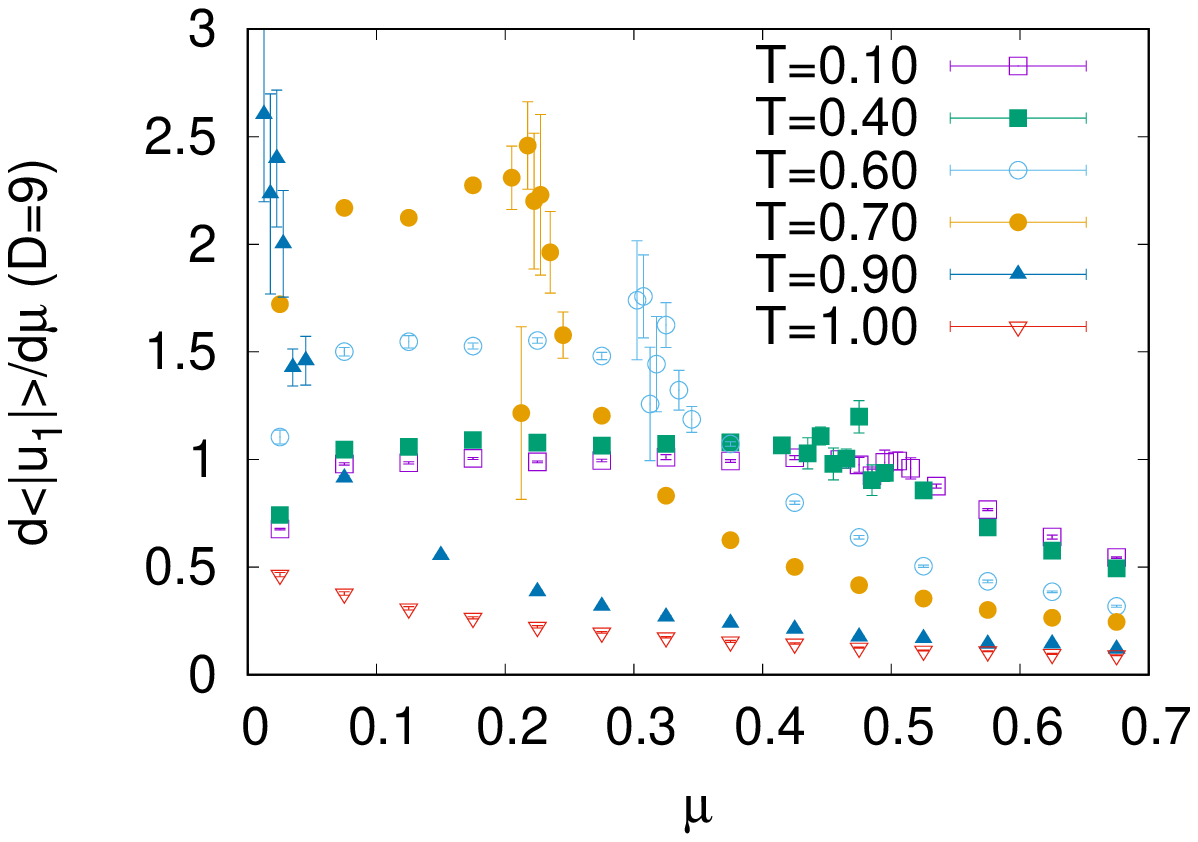}
\includegraphics[width=6.6cm]{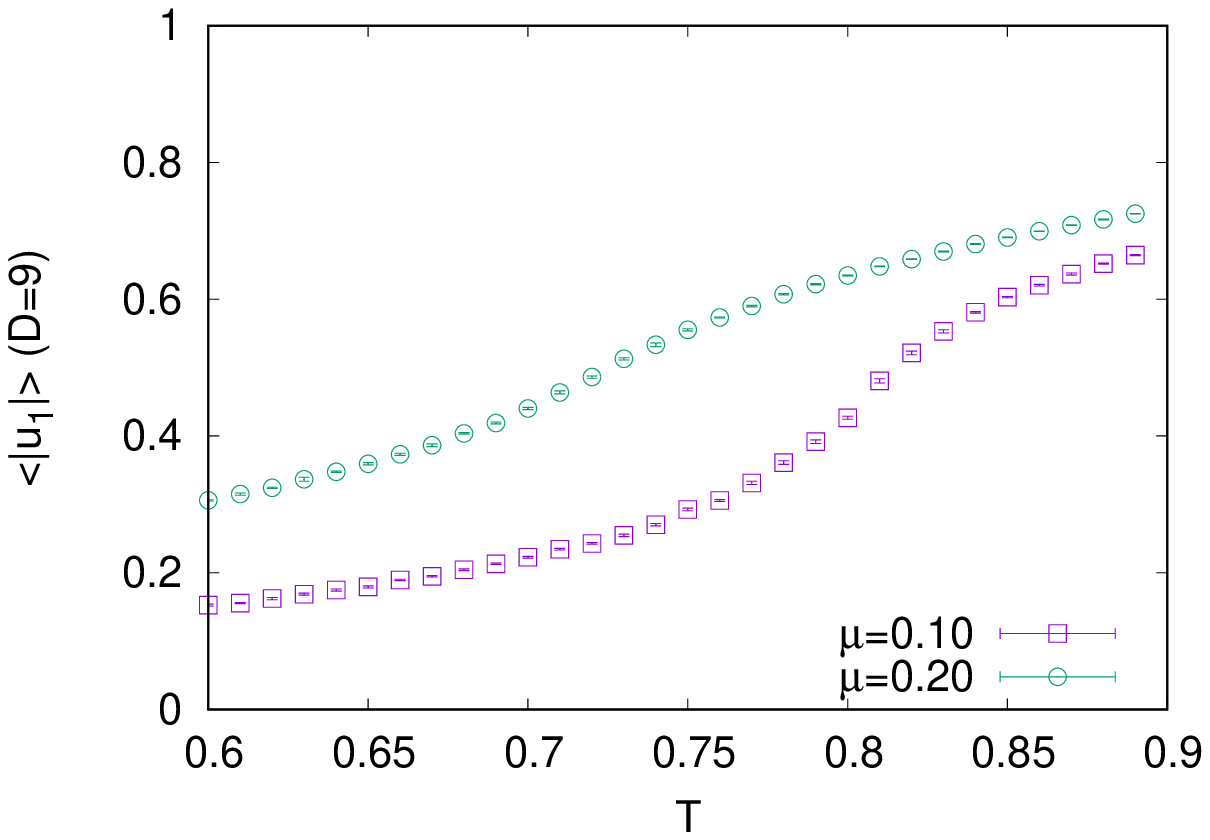}
\includegraphics[width=6.6cm]{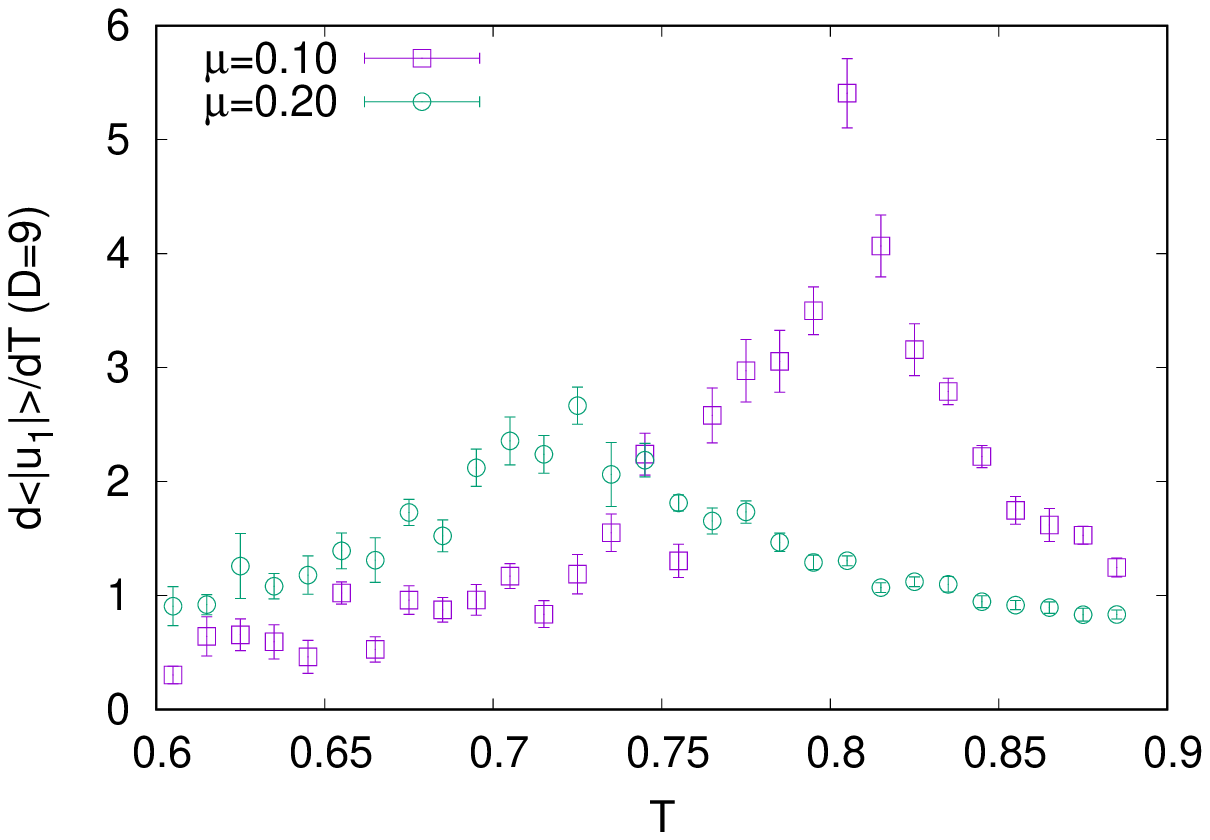}
%\vspace{4mm}
\caption{$\displaystyle \langle |u_{1}| \rangle$(left) and $\displaystyle \frac{d \langle |u_{1}| \rangle}{d \mu},\displaystyle \frac{d \langle |u_{1}| \rangle}{d T}$(right) for $D=9,N=48$ against $\mu$(top) and $T$(bottom).}
\label{un_data_d9}
\end{figure}

We numerically calculate the V.E.V.  $\displaystyle \langle |u_{1}| \rangle$ for $D=2,3,6,9$, $N=48$ and various $\displaystyle T = \frac{1}{\beta}$. We plot $\displaystyle \langle |u_{1}| \rangle$ against $\mu$ and $T$  in fig. \ref{un_data_d2}, \ref{un_data_d3}, \ref{un_data_d6}, \ref{un_data_d9}. $\displaystyle \langle |u_{1}| \rangle$ is monotonically increasing with respect to both $\mu$ and $T$. We find that at low temperature $T < T_{c0}$, the $\mu$-dependence bears resemblance to that of the unitary matrix model $\displaystyle S_{\textrm{g}}$, whose result is recapitulated in Appendix \ref{GWW_appendix}.

\begin{figure}[htbp]
%\vspace*{-15mm}
\centering % \begin{center}/\end{center} takes some additional vertical space
\includegraphics[width=7.4cm]{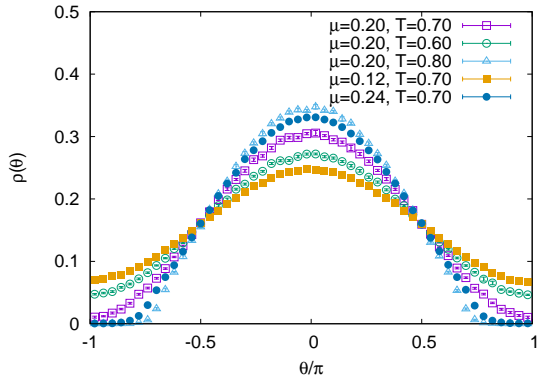}
\vspace{6mm}
\caption{\label{boson_dist} The eigenvalue distribution $\rho(\theta)$ for $D=3, N=48$ around the critical point $\displaystyle (\mu_{\textrm{c}}, T_{\textrm{c}}) \simeq (0.2, 0.7)$.}
%\vspace*{-15mm}
\end{figure}

In this case, we see that there is a signal of phase transition near the critical point $\displaystyle (\mu_c, T_c)$ at which $\langle |u_1| \rangle = 0.5$. The critical points are summarized in fig. \ref{critical_points} (p. \pageref{critical_points}) together with the case including the fermion, which we discuss later. At this point, the eigenvalue distribution 
\begin{eqnarray}
 \rho (\theta) = \frac{1}{N} \sum_{k=1}^{N} \langle \delta (\theta - \alpha_k) \rangle, \label{eig_distribution}
\end{eqnarray}
starts to develop a gap at the end $\theta =\pm \pi$. As a typical example, we plot in fig. \ref{boson_dist} the behavior of $\rho (\theta)$ for $D=3, N=48$ around the critical point $\displaystyle (\mu_{\textrm{c}}, T_{\textrm{c}}) \simeq (0.2, 0.7)$.

To see the nature of this transition, we plot $\displaystyle \frac{d\langle |u_1| \rangle}{d\mu}$ and  $\displaystyle \frac{d \langle |u_1| \rangle}{dT}$ in  fig. \ref{un_data_d2}, \ref{un_data_d3}, \ref{un_data_d6}, \ref{un_data_d9}, which are numerically approximated by the difference $\displaystyle \frac{ \langle |u_1| \rangle_{\mu=\mu_i}  - \langle |u_1| \rangle_{\mu=\mu_j}}{\mu_i - \mu_j}$ and $\displaystyle \frac{\langle |u_1| \rangle_{T=T_i}  - \langle |u_1| \rangle_{T=T_j}}{T_i - T_j}$ for the two neighboring points. We find that the derivatives $\displaystyle \frac{d \langle |u_1| \rangle}{d\mu}$ and  $\displaystyle \frac{d \langle |u_1| \rangle}{dT}$ are continuous but not smooth at the critical point $\displaystyle (\mu_c, T_c)$ (for example, $\displaystyle (\mu_c, T_c) \simeq (0.2, 0.7)$ at $D=2,3,6,9$) at the temperature region
\begin{eqnarray}
 T \leqq 0.9 \ (D=2,3), \ \ \ T \leqq 0.7 \ (D=6,9). \label{3rd_order} 
\end{eqnarray}
This suggests that the phase transition is possibly of third order, similarly to the  unitary matrix model $\displaystyle S_{\textrm{g}}$.\\

It is difficult to distinguish the order of the phase transition from $\displaystyle \frac{d\langle |u_1| \rangle}{d\mu}$ and  $\displaystyle \frac{d \langle |u_1| \rangle}{dT}$, especially for $0.7 <T<T_{c0}$. Instead, we try to fit our data with analytic functions in the regime $\mu < \mu_c$ and $\mu>\mu_c$ as \cite{0710_5873}
\begin{eqnarray}
 \langle | u_1 | \rangle =  \left\{ \begin{array}{ll} \displaystyle{q_1 \frac{\mu}{\mu_c} + r_1 \left( \frac{\mu}{\mu_c} \right)^2} & \displaystyle{\left( 0 \leqq \mu \leqq \mu_c \right)} \\ \displaystyle{1 - q_2 \left( \frac{\mu}{\mu_c} \right)^{-1} - r_2 \left( \frac{\mu}{\mu_c} \right)^{-2}} & \displaystyle{\left( \mu \geqq \mu_c \right)} \end{array} \right.  \label{ansatz_1}
\end{eqnarray}
This ansatz is based on the observation that, at large $N$, $\displaystyle \langle |u_1| \rangle =0$ at $\mu =0$ and that  $\displaystyle \langle |u_1| \rangle \to 1$ at $\mu \to +\infty$. And we also assume that $\displaystyle \langle | u_1| \rangle$ and $\displaystyle \frac{d \langle |u_1| \rangle}{d \mu}$ are continuous at $\mu =\mu_c$, which yields
\begin{eqnarray}
 r_1 = \frac{2 - 3 q_1 - q_2}{4}, \ \ r_2 = \frac{2 -  q_1 - 3 q_2}{4}. \label{ansatz_2}
\end{eqnarray} 
In this case, $\displaystyle \frac{d^2 \langle |u_1| \rangle}{d \mu^2}$ cannot be continuous. We obtain the coefficients $q_1, q_2$ from the fitting of the data at $0 \leqq \mu \leqq \mu_c$. Then using these values of $q_1, q_2$ and plugging it into eq. (\ref{ansatz_2}), we plot eq. (\ref{ansatz_1}) for the region $\mu \geqq \mu_{c}$. At $D=3, T=0.05$, as we see in fig. \ref{un_data_d3}, we obtain the coeffcients as
\begin{eqnarray}
 q_1 = 0.5046(10), \ \ q_2 = 0.5067(24). \label{ansatz_3}
\end{eqnarray}
This fitting turns out to work only at low temperature, but this bolsters that the system undergoes a third-order phase transition at low tempature.\\

\begin{figure}[htbp]
\centering % \begin{center}/\end{center} takes some additional vertical space
\includegraphics[width=7.4cm]{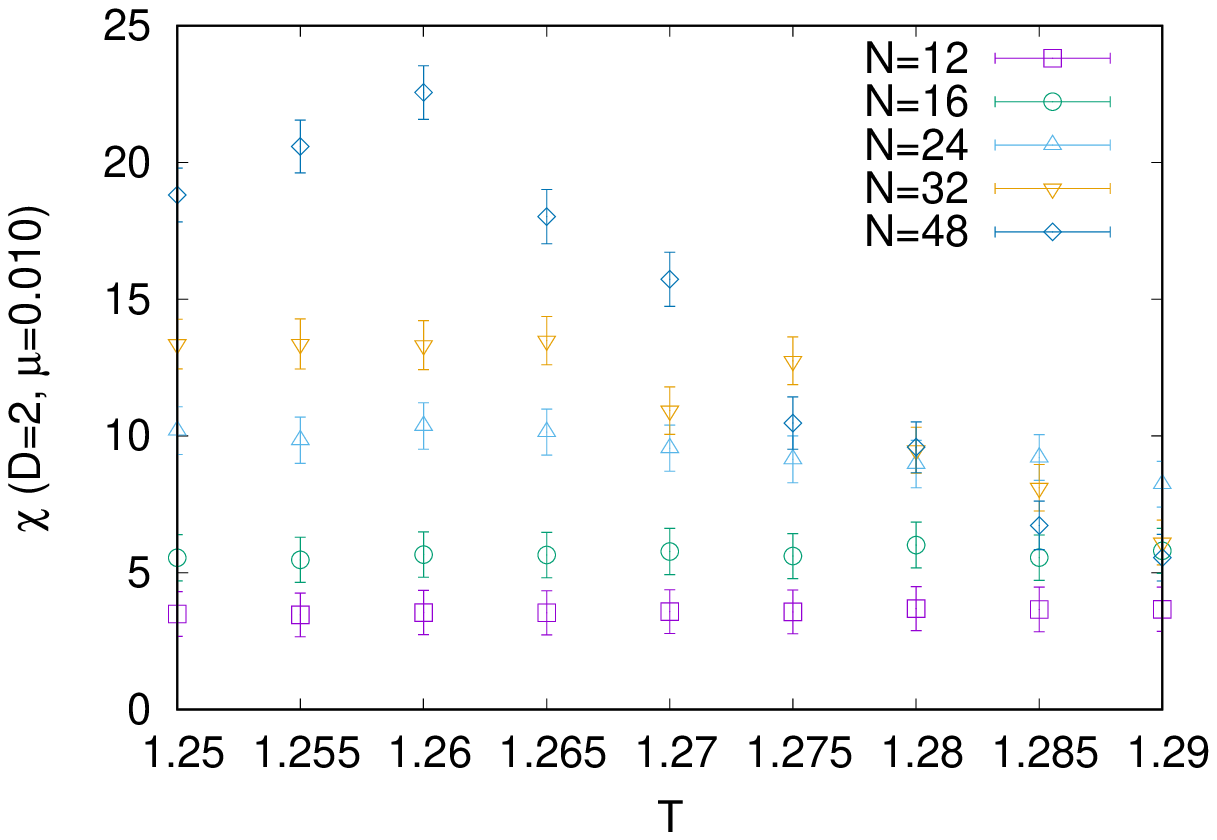}
\includegraphics[width=7.4cm]{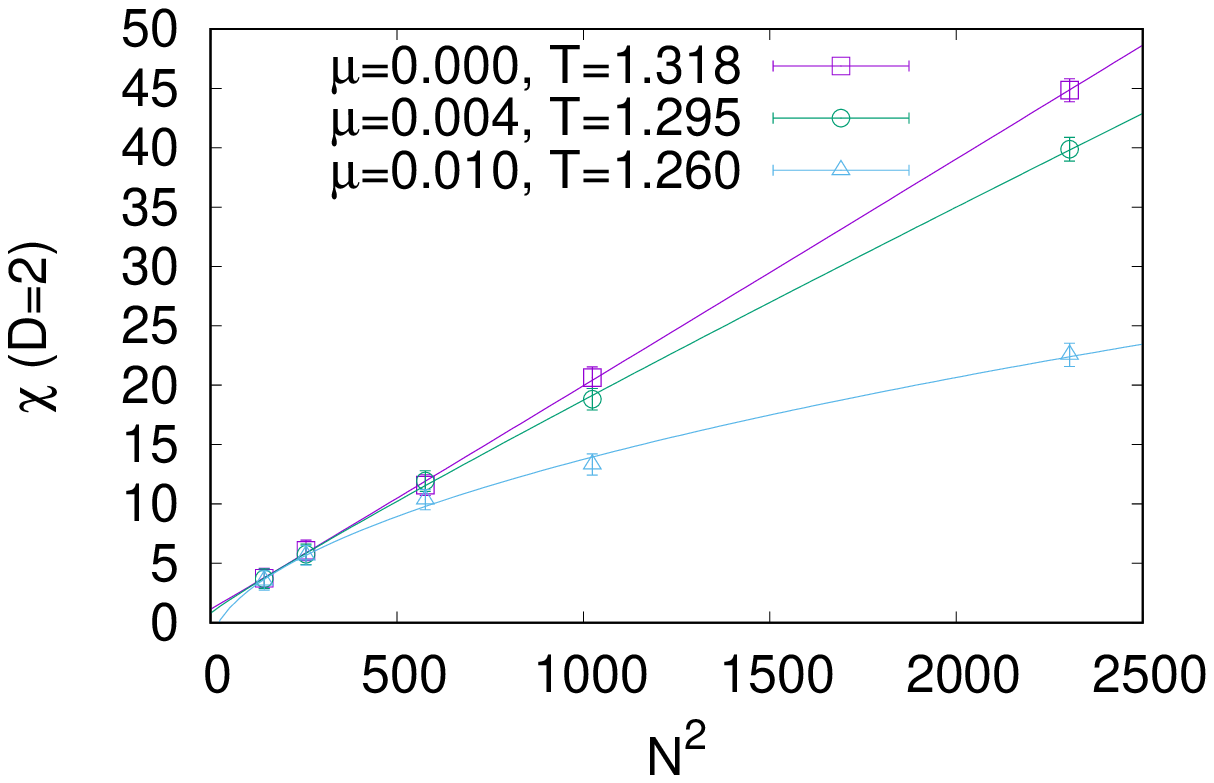}
\includegraphics[width=7.4cm]{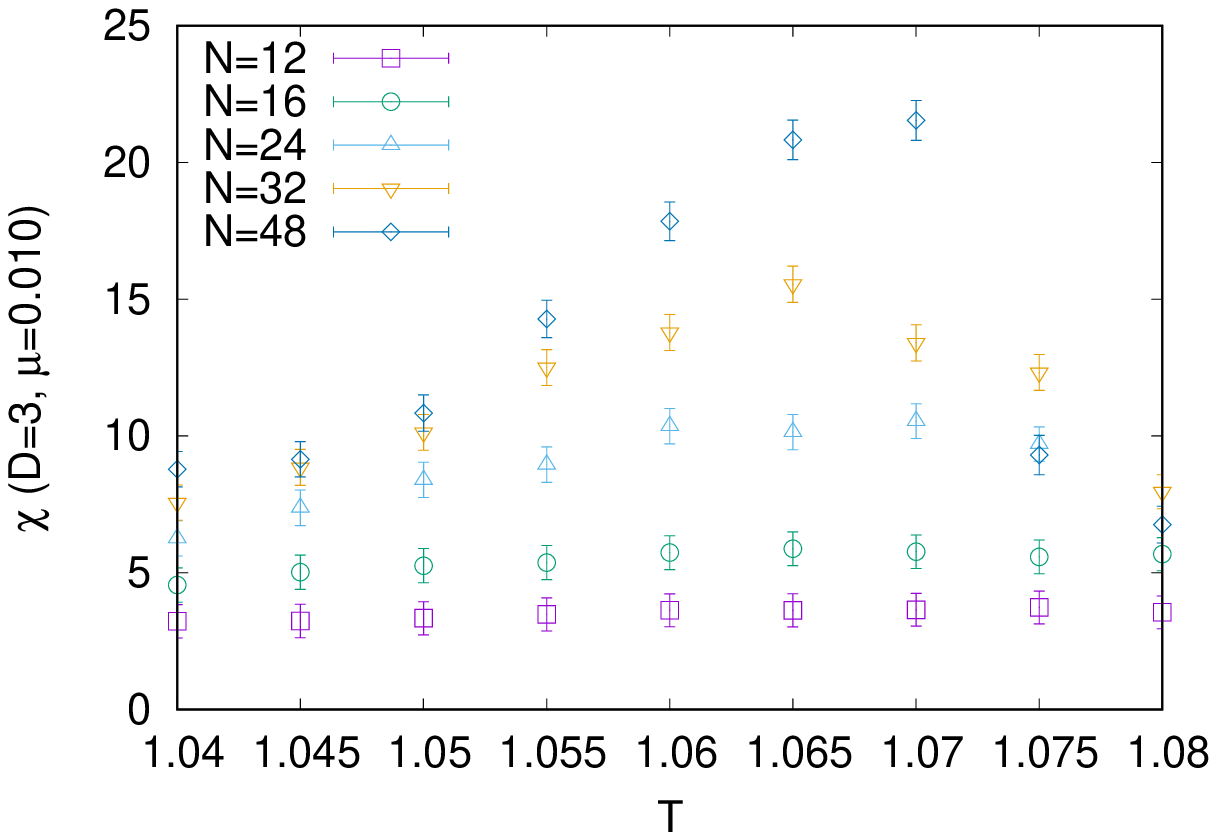}
\includegraphics[width=7.4cm]{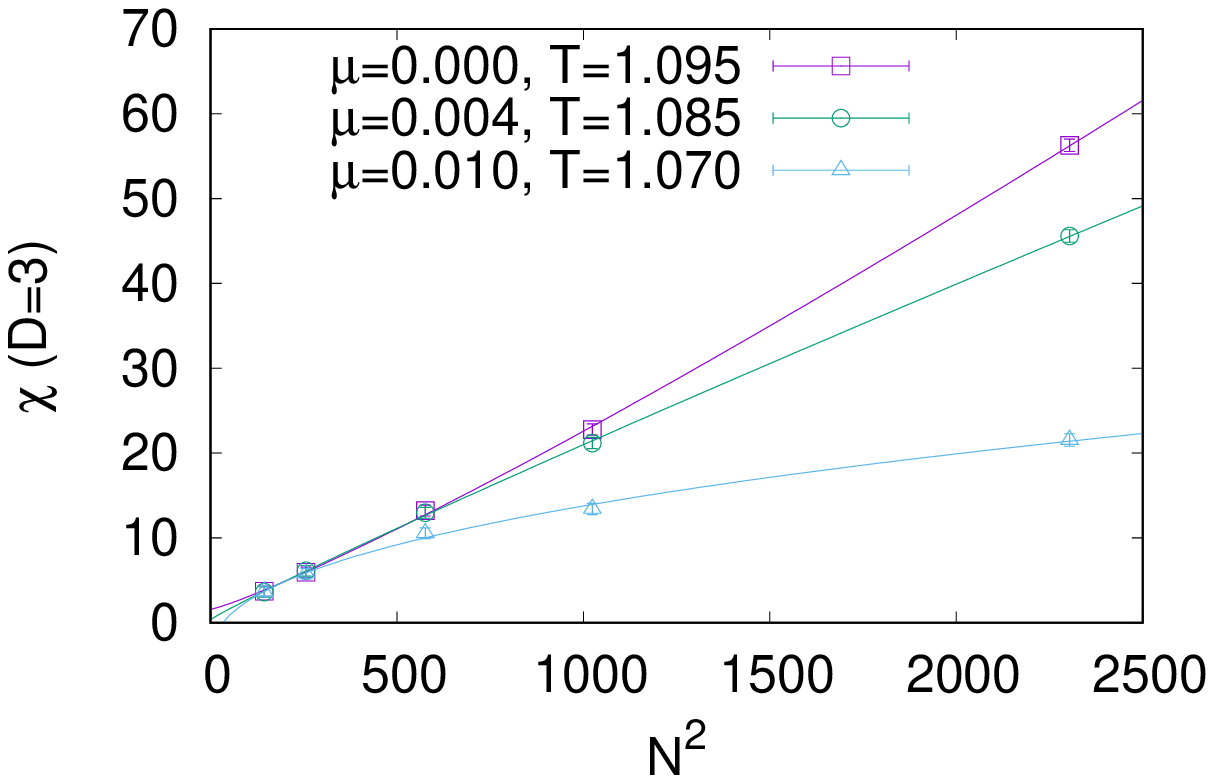}
\vspace{6mm}
\caption{Susceptibility $\chi$ for $D=2$ (top) and $D=3$ (bottom),  against $T$ for $\mu=0.010$ (left) and  against $N^2$ (right), the latter being fitted with $\displaystyle \chi = \gamma N^{2p}+c$.}
\label{suscep_result_fig}
\end{figure}

At $\mu=0$, the result of ref. \cite{1403_7764} suggests that the phase transition is of first order. In order to make comparison with this result, we consider the susceptibility 
\begin{eqnarray}
 \chi = N^2 \{ \langle |u_1|^2 \rangle - (\langle |u_1| \rangle)^2  \}. \label{susceptibitily}
\end{eqnarray}
For fixed $\mu$ we take a temperature $T$, where $\chi$ takes a maximum\footnote{Strictly speaking, this temperature slightly differs from the point where $\displaystyle \langle |u_1| \rangle = \frac{1}{2}$. For example. for $D=2, \mu=0.010$, $\chi$ takes a peak at $T=1.26$ while $\displaystyle \langle |u_1| \rangle = \frac{1}{2}$ at $T=1.28$.}. At this point we fit $\chi$  as
\begin{eqnarray}
 \chi = \gamma N^{2p} + c \label{suscep_fit}
\end{eqnarray}
with the fitting parameters ($\gamma,p,c$). If the power $p$ is 1, this suggests that the phase transition is of first order \cite{susceptibility_cite}. We plot $\chi$ against $N^2$ for $D=2,3$ for brevity in fig. \ref{suscep_result_fig}. The power $p$ is obtained in table \ref{suscep_result_table}. This suggests that the phase transition becomes no longer of first order even for small positive $\mu$. 

\begin{table}[htbp]
\begin{center} 
\begin{tabular}{|l||l|l|l||l|l|l|}
\hline
$D$     & 2          & 2         & 2          & 3         & 3         & 3       \\ \hline \hline
$\mu_c$ & $0.000$    & $0.004$   & $0.010$    & $0.000$   & $0.004$   & $0.010$  \\ \hline
$T_c$   & $1.318$    & $1.295$   & $1.260$    & $1.095$   & $1.085$   & $1.070$  \\ \hline
$p$     & $1.01(3)$  & $0.93(4)$ & $0.52(10)$ & $1.14(4)$ & $0.94(3)$ & $0.42(10)$ \\ \hline
\end{tabular}
\caption{The power $p$ in the fitting of the susceptibility $\displaystyle \chi = \gamma N^{2p}+c$.}
\label{suscep_result_table}
\end{center}
\end{table}

At high temperature $T>T_{c0}$, there is no such phase transition, as studied in ref. \cite{0710_5873}. We have studied in fig. \ref{un_data_d2}, \ref{un_data_d3}, \ref{un_data_d6}, \ref{un_data_d9} the behavior at 
\begin{eqnarray}
 T=1.4 (D=2), \ \ T=1.2 (D=3), \ \ T=1.0 (D=6,9). \ 
\end{eqnarray}
In these cases, $\displaystyle \langle | u_1 | \rangle$ is already over 0.5 at $\mu=0$. $\displaystyle \langle | u_1 | \rangle$  and $\displaystyle \frac{d\langle |u_1| \rangle}{d\mu}$ are smooth function with respect to $\mu$. In this case, the system is always in the deconfined phase. 

\section{Phase Transitions in a Fermionic Matrix Model} \label{sec_fermionic_model}
\subsection{Adding Fermions at Finite Temperature}
\noindent In this section, we show preliminary results on the case of adding fermions. We study the following action for $D=3$.
\begin{eqnarray}
 S &=& S_{\textrm{b}} + S_{\textrm{g}}  + S_{\textrm{f}}, \textrm{ where } \label{ferm_action} \\
 S_{\textrm{b}} &=& \frac{1}{g^2} \int^{\beta}_{0}  \textrm{tr} \left\{ \frac{1}{2} \sum_{\mu=1}^{3} (D_t X_{\mu} (t))^2 - \frac{1}{4} \sum_{\mu,\nu=1}^{3}  [X_{\mu} (t), X_{\nu} (t) ]^2 \right\} dt, \ \ S_{\textrm{g}} = N \mu (\textrm{tr} U + \textrm{tr} U^{\dagger}), \nonumber \\ \label{bosonic_action2} \\
 S_{\textrm{f}} &=& \frac{1}{g^2} \int^{\beta}_{0} \textrm{tr} \left\{  \sum_{\alpha=1}^{2} {\bar \psi}_{\alpha} (t) D_t \psi_{\alpha} (t) -  \sum_{\mu=1}^{3}  \sum_{\alpha,\eta=1}^{2} {\bar \psi}_{\alpha} (t) (\sigma_{\mu})_{\alpha \eta} [X_{\mu} (t), \psi_{\eta} (t)] \right\} dt. \label{ferm_action2}
\end{eqnarray}
Here again, $\displaystyle \beta = \frac{1}{T}$ is the inverse temperature, and we take the unit $g^2 N=1$. For $D=3$, $\displaystyle S_{\textrm{b}}$ and $\displaystyle S_{\textrm{g}}$ are the same as eq. (\ref{bosonic_action}) and (\ref{chemical_pot}), respectively. $\sigma_{\mu}$ are $2 \times 2$ Pauli matrices
\begin{eqnarray}
 \sigma_1 = \left( \begin{array}{cc} 0 & 1 \\ 1 & 0  \end{array} \right), \ \ \sigma_2 = \left( \begin{array}{cc} 0 & -i \\ i & 0  \end{array} \right), \ \ \sigma_3 = \left( \begin{array}{cc} 1 & 0 \\ 0 & -1  \end{array} \right). \label{pauli_gamma}
\end{eqnarray}
$\psi(t)$ are $N\times N$ traceless matrices with complex Grassmann entries. While we have a periodic boundary condition (\ref{periodic_boundary}) for $A(t)$ and $X_{\mu}(t)$, we impose an anti-periodic boundary condition on $\psi(t)$
\begin{eqnarray}
 \psi( t + \beta) = - \psi(t). \label{anti_periodic_boundary}
\end{eqnarray}
Here again, we take a static diagonal gauge (\ref{static_diagonal}), where $\alpha_k$ are chosen to satisfy the constraint \cite{1603_00538}
\begin{eqnarray}
 - \pi \leqq \alpha_k < \pi. \label{large_gauge}
\end{eqnarray}
We add a gauge-fixing term $\displaystyle S_{\textrm{g.f.}} = - \sum_{k,l=1, k \neq l}^{N} \log \left| \sin \frac{\alpha_k - \alpha_l}{2} \right|$, which is the same as eq. (\ref{gauge_fixing}).
For $\mu=0$ (without the term $\displaystyle S_{\textrm{g}}$), this model is a dimensional reduction of the four-dimensional ${\cal N}=1$ U$(N)$ super-Yang-Mills theory to one dimension. This supersymmetric model, as well as the $D=9$ version, has been intensively studied in refs. \cite{0706_1647,0707_4454,0901_4073,1012_2913,1108_5153,1311_5607,1603_00538}.

Here, the term $\displaystyle S_{\textrm{g}}$ breaks the supersymmetry, as well as the invariance under (\ref{a_trans}). In putting this action on a computer, we make a Fourier expansion \cite{0706_1647,0707_4454}, instead of the lattice regularization as in the bosonic case $\displaystyle \left( \omega = \frac{2\pi}{\beta} \right)$:
\begin{eqnarray}
 \hspace*{-4mm} X_{\mu}^{kl} (t) = \sum_{n= - \Lambda}^{\Lambda} {\tilde X}_{\mu,n}^{kl} e^{i \omega n t}, \ \ \psi_{\alpha}^{kl} (t) = \sum_{r= - \Lambda + \frac{1}{2}}^{\Lambda- \frac{1}{2}} {\tilde \psi}_{\alpha,r}^{kl} e^{i \omega r t}, \ \ {\bar \psi}_{\alpha}^{kl} (t) = \sum_{r= - \Lambda + \frac{1}{2}}^{\Lambda- \frac{1}{2}} {\tilde {\bar \psi}}_{\alpha,-r}^{kl} e^{i \omega r t}. \label{fourier_expansion}
\end{eqnarray}
The indices $n$ and $r$ take integer and half-integer values, respectively. At finite $\Lambda$, the supersymmetry is broken due to the difference of the degrees of freedom between ${\tilde X}_{\mu,n}$ and ${\tilde \psi}_{\alpha,r}$. From $\displaystyle \int^{\beta}_{0} e^{i \omega n t} dt = \beta \delta_{n,0}$, we eventually simulate the action
\begin{eqnarray}
 S_{\textrm{Fourier}} &=& S_{\textrm{B,Fourier}} + S_{\textrm{F,Fourier}} +  2 N \mu \sum_{k=1}^{N} \cos \alpha_k - \sum_{k,l=1, k \neq l}^{N} \log \left| \sin \frac{\alpha_k - \alpha_l}{2} \right|, \textrm{ where } \nonumber \\ \label{our_ferm_action_fourier} \\
 S_{\textrm{B,Fourier}} &=& N \beta \left\{ \frac{1}{2} \sum_{n = - \Lambda}^{\Lambda} \left\{ n \omega  - \frac{\alpha_k - \alpha_l}{\beta} \right\}^2 {\tilde X}_{\mu,n}^{kl} {\tilde X}_{\mu,-n}^{lk} - \frac{1}{4} \textrm{tr} \left( [{\tilde X}_{\mu}, {\tilde X}_{\nu}]^2 \right)_0 \right\}, \label{boson_fourier2} \\
 S_{\textrm{F,Fourier}} &=& N \beta \sum_{r = - \Lambda+ \frac{1}{2}}^{\Lambda- \frac{1}{2}} \left\{ i \left\{ r \omega  - \frac{\alpha_k - \alpha_l}{\beta} \right\} {\tilde {\bar \psi}}_{\alpha,r}^{lk} {\tilde \psi}_{\alpha,r}^{kl} - (\sigma_{\mu})_{\alpha \eta} \textrm{tr} \left\{ [{\tilde {\bar \psi}}_{\alpha,r} \left( [ {\tilde X}_{\mu}, {\tilde \psi}_{\eta} ] \right)_r \right\} \right\}. \nonumber \\ \label{ferm_fourier2}
\end{eqnarray}
We have introduced a short-hand notation
\begin{eqnarray}
 \left( f^{(1)} \cdots f^{(p)} \right)_q = \sum_{k_1+\cdots+k_p=q} f^{(1)}_{k_1} \cdots f^{(p)}_{k_p}, \label{fourier_short}
\end{eqnarray}
where the indices $k_i$ ($i=1,2,\cdots,p$) run over $k_i = -\Lambda, -\Lambda+1, \cdots, \Lambda$ for $\displaystyle {\tilde X}_{\mu}$, and $\displaystyle k_i = -\Lambda + \frac{1}{2}, -\Lambda + \frac{3}{2}, \cdots, \Lambda- \frac{1}{2}$ for $\displaystyle {\tilde \psi}_{\alpha}$, respectively. The continuum limit is realized by taking the $\Lambda \to \infty$ limit.
Integrating out ${\bar \psi}, \psi$ in the action $S_{\textrm{F,Fourier}}$ yields an $N_0 \times N_0$ matrix ${\cal M}$ with $N_0 = 2 \times 2 \Lambda \times (N^2-1)$. In $D=3$, $\det {\cal M}$ is real and there is no sign problem.\footnote{We comment on the differences in simulating the $D=9$ case.  $\psi$ comes from the 16-component Majorana-Weyl fermion. The Pauli matrices $\sigma_{\mu}$ are replaced by the $16 \times 16$ Gamma matrices satisfying the Euclidean Clifford algebra $\{ \gamma_{\mu}, \gamma_{\nu} \} = 2 \delta_{\mu \nu}$. 
An example of such matrices is
\begin{eqnarray}
& & \gamma_1 =  \sigma_2 \otimes \sigma_2 \otimes \sigma_2 \otimes \sigma_2, \ \ \gamma_2 =  \sigma_2 \otimes \sigma_2 \otimes {\bf 1} \otimes \sigma_1, \ \ \gamma_3 =  \sigma_2 \otimes \sigma_2 \otimes {\bf 1} \otimes \sigma_3, \nonumber \\
& & \gamma_4 = \sigma_2 \otimes \sigma_1 \otimes \sigma_2 \otimes {\bf 1}, \ \ \gamma_5 = \sigma_2 \otimes \sigma_3 \otimes \sigma_2 \otimes {\bf 1}, \ \  \gamma_6 = \sigma_2 \otimes {\bf 1} \otimes \sigma_1 \otimes \sigma_2, \nonumber \\
& & \gamma_7 = \sigma_2 \otimes {\bf 1} \otimes \sigma_3 \otimes \sigma_2, \ \ \gamma_8 = \sigma_1 \otimes {\bf 1} \otimes {\bf 1} \otimes {\bf 1}, \ \ \gamma_9 = \sigma_3 \otimes {\bf 1} \otimes {\bf 1} \otimes {\bf 1}.  \label{9d_gamma}
\end{eqnarray}
The term (\ref{ferm_action2}), and its Fourier expansion (\ref{ferm_fourier2}) are also replaced by
\begin{eqnarray}
 S_{\textrm{f}} &=& N \int^{\beta}_{0} \textrm{tr} \left\{ \sum_{\alpha=1}^{16} \psi_{\alpha} (t) D_t \psi_{\alpha} (t) - \sum_{\mu=1}^{9}  \sum_{\alpha,\eta=1}^{16} \psi_{\alpha} (t) (\gamma_{\mu})_{\alpha \eta} [X_{\mu} (t), \psi_{\eta} (t)] \right\}, \label{ferm_action9} \\
 S_{\textrm{F,Fourier}} &=& N \beta \sum_{r = - \Lambda+ \frac{1}{2}}^{\Lambda- \frac{1}{2}} \left\{ i \left\{ r \omega  - \frac{\alpha_k - \alpha_l}{\beta} \right\} {\tilde \psi}_{\alpha,-r}^{lk} {\tilde \psi}_{\alpha,r}^{kl} - (\gamma_{\mu})_{\alpha \eta} \textrm{tr} \left\{ [{\tilde \psi}_{\alpha,-r} \left( [ {\tilde X}_{\mu}, {\tilde \psi}_{\eta} ] \right)_r \right\} \right\}.  \label{ferm_fourier2_9}
\end{eqnarray}
After integrating out $\psi$, we have an $N_0 \times N_0$ matrix ${\cal M}$ with $N_0 = 16 \times 2 \Lambda \times (N^2-1)$. Its Pfaffian $\displaystyle \textrm{Pf} {\cal M}$ is complex in general. However, as pointed out in ref. \cite{1603_00538}, its complex phase can be neglected at sufficiently high or low temperature.}
We employ a rational Hybrid Monte Carlo (RHMC) algorithm \cite{HMC_ref,hep-lat9809092} with a multi-mass solver \cite{hep-lat9612014}, whose details we delegate to Appendix B of ref. \cite{1108_5153}\footnote{In RHMC, we introduce the pseudofermion, as in eq. (B.22) of ref. \cite{1108_5153}. We update the pseudofermion via heatbath algorithm, instead of solving the Hamiltonian equation as in eq. (B.25) of ref. \cite{1108_5153}.} (its pedagogic review is found in Chapter II6,II7 of ref. \cite{1506_02567}).

In ref. \cite{0707_4454,1012_2913}, it was pointed out that at $\mu=0$ (without the chemical potential term $S_{\textrm{g}}$) the V.E.V. of the Polyakov loop behaves as
\begin{eqnarray}
 \langle | u_1 | \rangle = a_0 e^{-a_1/T}, \label{no_CD_SUSY}
\end{eqnarray}
with some constants $a_0,a_1$ at $T \geqq 0.4$.
% and hence that there is no confinement/deconfinement phase transition.

\subsection{Results for the Fermionic Matrix Model}
\begin{figure}[htbp]
\centering % \begin{center}/\end{center} takes some additional vertical space
\includegraphics[width=7.4cm]{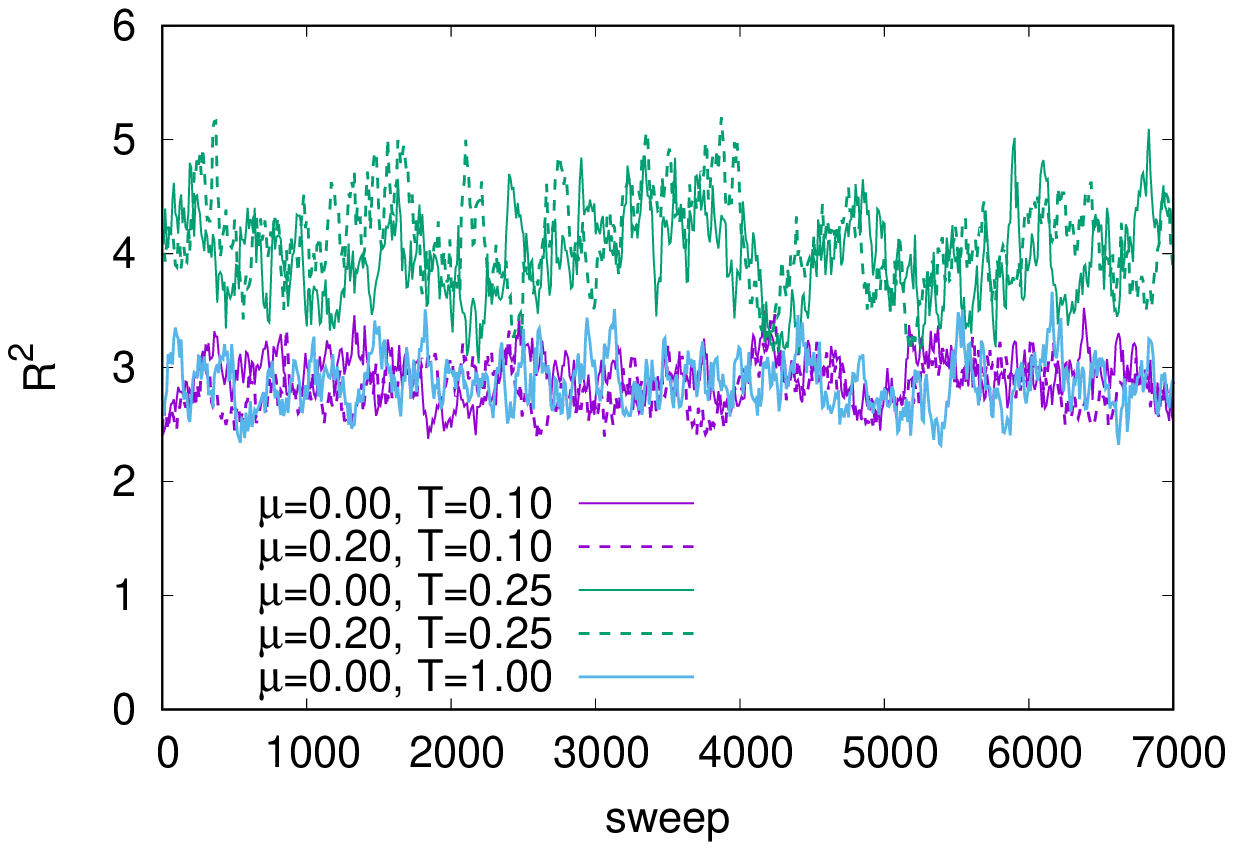}
\includegraphics[width=7.4cm]{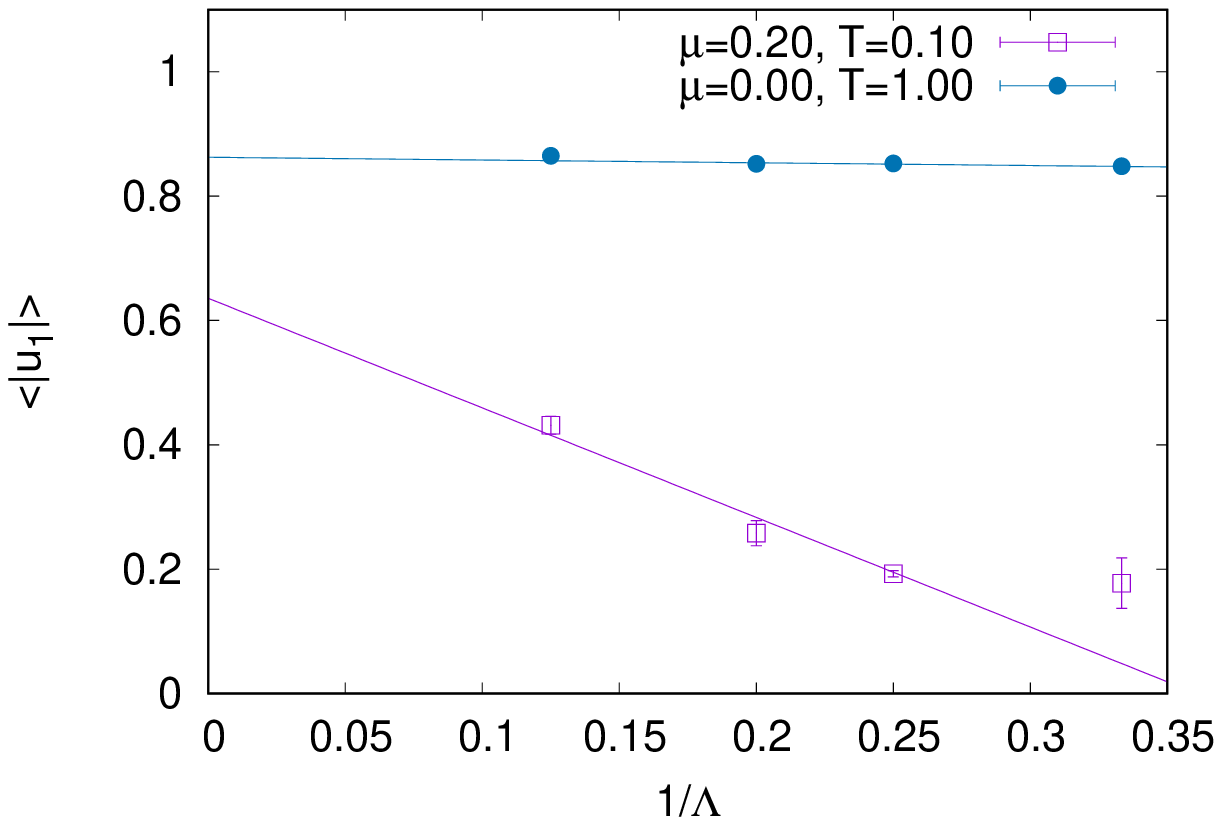}
\vspace{6mm}
\caption{\label{SUSY_result2} (Left) The history of $R^2$ for $D=3, N=16, \Lambda=3$. (Right) An example of the large-$\Lambda$ extrapolation for $D=3, N=16$.}
\end{figure}

The V.E.V. $\langle | u_1 | \rangle$ is below $0.5$ only at low temperature \cite{1012_2913}. This leads us to study the case $T=0.10, \ 0.12, \ 0.15, \ 0.20, \ 0.25$, as well as the high-temperature case $T=1.00$ case, for $D=3, N=16$. 
In fig.  \ref{SUSY_result2} (Left), we plot the history of 
\begin{eqnarray}
 R^2 = \frac{1}{N \beta} \int^{\beta}_{0} \textrm{tr} \{ X_{\mu} (t) \}^2 dt \label{r_def}
\end{eqnarray}
for typical values of $(\mu, T)$ at $D=3, N=16, \Lambda=3$. This suggests that there is no instability coming from the flat direction \cite{0707_4454}.
For low temperature, the result is affected by a finite-$\Lambda$ effect and we make a large-$\Lambda$ extrapolation by fitting the observables for $\Lambda=3,4,5,8$ as
\begin{eqnarray}
 \langle |u_1| \rangle = b_0 + \frac{b_1}{\Lambda}. \label{large-lambda}
\end{eqnarray}
An example of this extrapolation is given in fig. \ref{SUSY_result2} (Right). At high temperature $T=1.00$,  the inverse temperature $\displaystyle \frac{1}{T}$ is small enough that the $\Lambda$-depndence is negligible, which leads us to omit this extrapolation and put the result for $\Lambda=8$ for brevity. 

\begin{figure}[htbp]
\centering % \begin{center}/\end{center} takes some additional vertical space
\includegraphics[width=7.4cm]{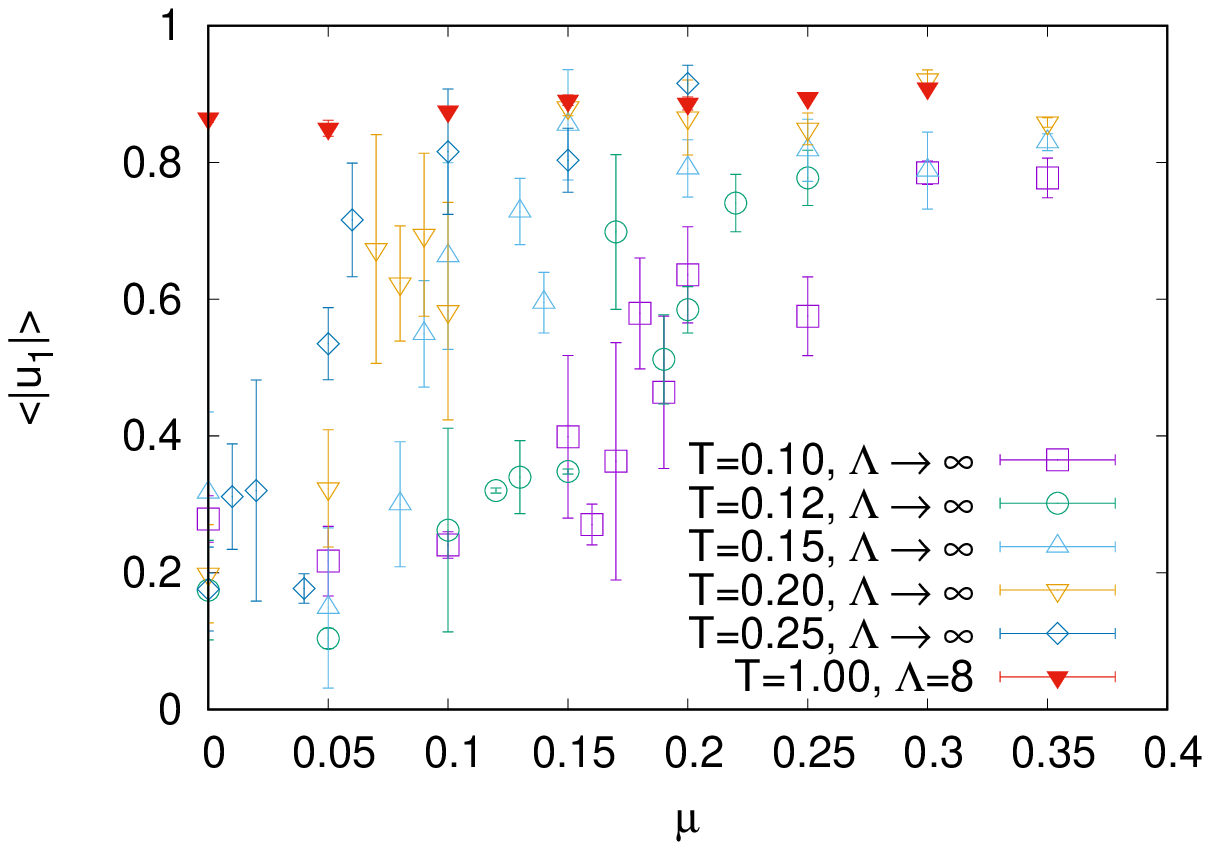}
\includegraphics[width=7.4cm]{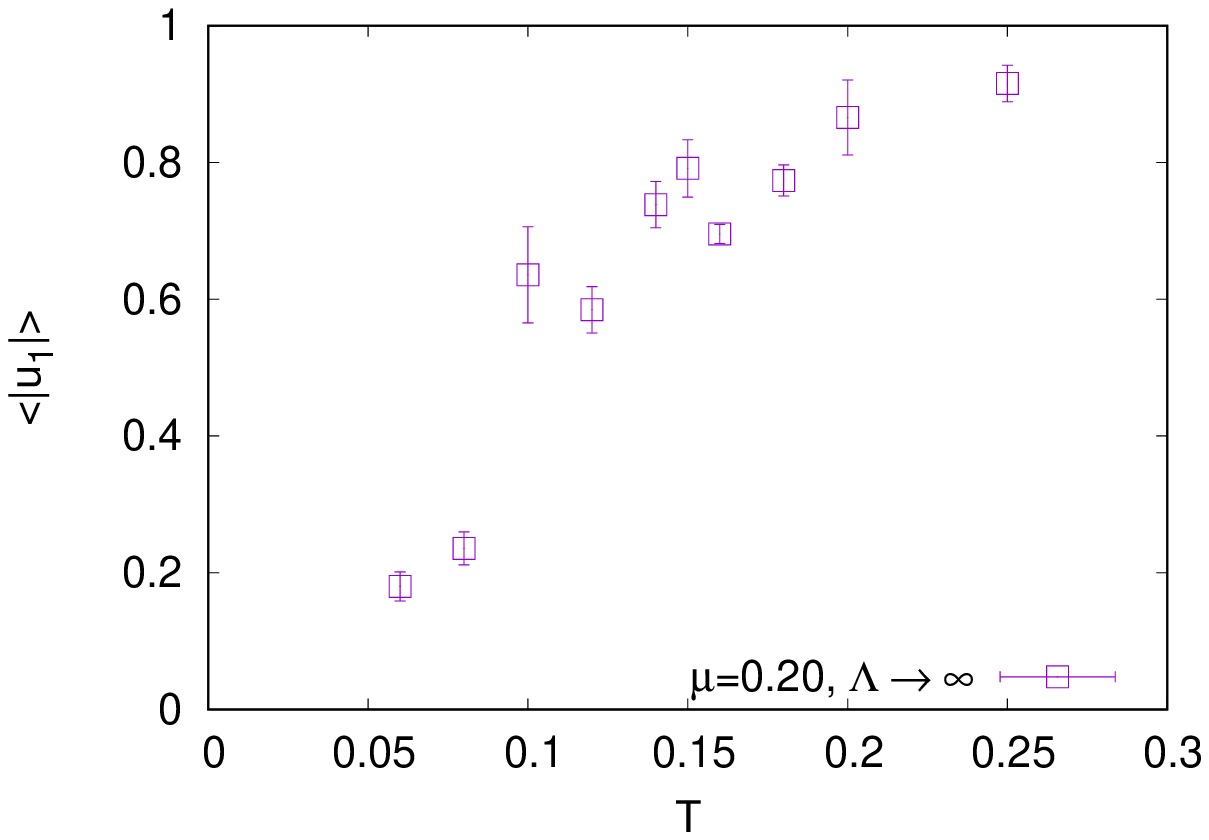}
\vspace{6mm}
\caption{\label{SUSY_result} The V.E.V. $\langle |u_{1} | \rangle$ for $D=3, N=16$. For the high temperature $T=1.00$, we put the result of $\Lambda=8$ without large-$\Lambda$ extrapolation. }
\end{figure}

\begin{figure}[htbp]
\centering % \begin{center}/\end{center} takes some additional vertical space
\includegraphics[width=7.4cm]{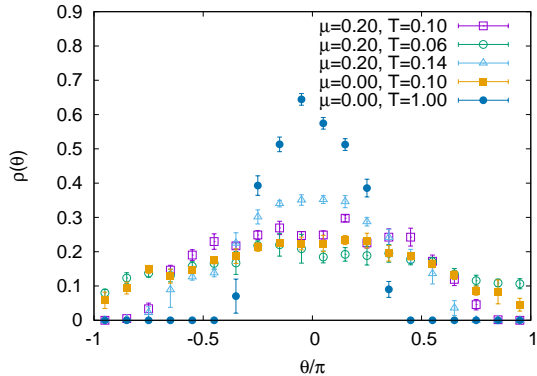}
\vspace{6mm}
\caption{\label{SUSY_result3} The eigenvalue distribution $\rho(\theta)$ for $D=3, N=16, \Lambda=8$.}
\end{figure}

The V.E.V. $\langle |u_1| \rangle$ is summarized in fig. \ref{SUSY_result}. 
The eigenvalue distribution (\ref{eig_distribution}) is summarized for $D=3$ in fig. \ref{SUSY_result3}. At $\mu=0$ (without the chemical potential term $S_{\textrm{g}}$), the eigenvalue distribution (\ref{eig_distribution}) is gapped at $T=1.00$ ($\displaystyle \langle |u_1| \rangle > 0.5$), and ungapped at $T=0.10$  ($\displaystyle \langle |u_1| \rangle < 0.5$). As we change the coefficient $\mu$, we encounter a point $\displaystyle (\mu_{c}, T_{c})$, at which $\displaystyle \langle |u_1| \rangle =\frac{1}{2}$, such as $\displaystyle (\mu_{c}, T_{c}) = (0.10,0.20)$. At this point the eigenvalue distribution (\ref{eig_distribution}) starts to develop a gap at the ends $\theta = \pm \pi$.
%\footnote{Strictly speaking, in fig. \ref{SUSY_result2} (right) we plot the eigenvalue distribution (\ref{eig_distribution}) for $\Lambda = 8$ without large-$\Lambda$ extrapolation for brevity.}
At higher $T$ or $\mu$ the eigenvalue distribution (\ref{eig_distribution}) becomes gapped, and at lower $T$ or $\mu$ it becomes ungapped.
This suggests a possible phase transition between the gapped and ungapped phase at the points $\displaystyle (\mu_{c}, T_{c})$, {\it including} the $\mu_{c}=0$ case (in the absence of the chemical potential term).

\section{Phase Diagram of the Fermionic Model} \label{sec_phase_diag}
\begin{figure}[htbp]
\centering % \begin{center}/\end{center} takes some additional vertical space
\includegraphics[width=7.4cm]{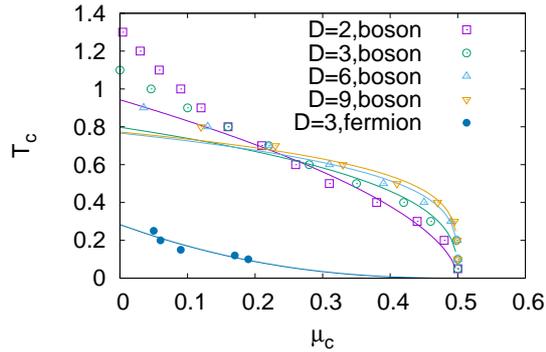}
\vspace{6mm}
\caption{\label{critical_points} The point $(\mu_{\textrm{c}}, T_{\textrm{c}})$ at which $\displaystyle \langle |u_1| \rangle = 0.5$. The result of the bosonic action (\ref{our_bosonic_action}) for $D=2,3,6,9$, $N=48$ is indicated by "boson". The result of the action including the fermion (\ref{ferm_action}) for $D=3$, $N=16$ is indicated by "fermion".}
\end{figure}

We have studied the bosonic action (\ref{our_bosonic_action}) and the fermionic action (\ref{ferm_action}). 
In the following, $(\mu_c, T_c)$ are the points at which $\displaystyle \langle |u_1| \rangle =\frac{1}{2}$. 
In the bosonic action (\ref{our_bosonic_action}) we have found that these are the critical points of the third-order GWW-type phase transition. 
It is interesting that in the fermionic action (\ref{ferm_action}) as well, the eigenvalue distribution  (\ref{eig_distribution}) suggests a phase transition at $(\mu_c, T_c)$. The points $(\mu_c, T_c)$ are summarized in fig. \ref{critical_points}.
In the absence of the scalar fields $X_{\mu} (t)$, the GWW third-order phase transition occurs at $\displaystyle \mu = 0.5$, as is presented in Appendix \ref{GWW_appendix}. This leads us to fit the points $\displaystyle (\mu_{\textrm{c}}, T_{\textrm{c}})$ by the curve 
\begin{eqnarray}
 T_{\textrm{c}} = a (0.5 - \mu_{\textrm{c}})^b \label{power_cr}
\end{eqnarray}
with respect to the bosonic action (\ref{our_bosonic_action}) at $D=2,3,6,9$, and the fermionic action (\ref{ferm_action}) at $D=3$. The $D$-dependence of the power $b$ is obtained in table \ref{Tab-results}.

\begin{table}[htbp]
\begin{center} 
\begin{tabular}{|c||l|l|l|l||l|}
\hline
$D$ & 2(boson) & 3(boson) & 6(boson) & 9(boson)  & 3(fermion) \\ \hline \hline
$a$ & 1.36(12) & 1.01(15) & 0.91(9) & 0.90(8) & 1.39(72) \\ \hline
$b$ & 0.55(6)  & 0.34(7)  & 0.25(4) & 0.23(4) & 2.30(59) \\ \hline
\end{tabular}
\caption{The coefficients in the fitting (\ref{power_cr}) for the bosonic model (\ref{our_bosonic_action}) at $D=2,3,6,9$, and the fermionic model (\ref{ferm_action}) at $D=3$.}
\label{Tab-results}
\end{center}
\end{table}

The bosonic result can be compared with the phase diagram in figure 1 of ref. \cite{1110_0696}. The power behavior (\ref{power_cr}) for $0<b<1$ is consistent with figure 1 of ref. \cite{1110_0696} in that this curve is convex upward. On the other hand, the saw-tooth features at low temperature and $\mu>0.5$ could not be observed in the bosonic action (\ref{our_bosonic_action}), since $\langle |u_1| \rangle$ simply increases monotonically with respect to $\mu$ at low temperature, as is indicated in fig. \ref{un_data_d2}, \ref{un_data_d3}, \ref{un_data_d6}, \ref{un_data_d9}.

In the fermionic action (\ref{ferm_action}), we see a tangible difference in the value of the critical point from the bosonic case for $D=3$. In ref. \cite{1012_2913}, it was shown that at $\mu=0$ $\langle |u_1| \rangle$ behaves as $\displaystyle \langle |u_1| \rangle = a_0 e^{-a_1/T}$, with $a_0=1.03(1)$ and $a_1=0.19(1)$, which implies that $\displaystyle \langle |u_1| \rangle =\frac{1}{2}$ at $\displaystyle T = \frac{-a_1}{\log \left( \frac{a_0}{2} \right)} \simeq 0.28$. Our fitting (\ref{power_cr}) suggests that at $\mu=0$ the temperature at which $\displaystyle \langle |u_1| \rangle =\frac{1}{2}$ is $\displaystyle T_c = a \times 0.5^b = 1.39 \times 0.5^{2.30} \simeq 0.28$, which is consistent with the result of ref. \cite{1012_2913}.

\section{Conclusions} \label{sec_conclusion}
In this paper, we have studied the phase transition of the finite-temperature matrix quantum mechanics with a chemical potential term using Monte Carlo simulation. In the bosonic case, we have observed a GWW-type third-order phase transition at large $N$, except for very small $\mu$ (the coefficient of the term $\displaystyle S_{\textrm{g}} = N \mu (\textrm{tr} U + \textrm{tr} U^{\dagger})$). In that case, we have numerically shown that the derivatives $\displaystyle \frac{d \langle |u_1| \rangle}{d \mu}$ and  $\displaystyle \frac{d \langle |u_1| \rangle}{d T}$ are continuous but not smooth at the critical point. This behavior is akin to that of the unitary matrix model $\displaystyle S_{\textrm{g}}$. 
We have also studied the matrix model with fermionic degrees of freedom, using the non-lattice simulation via Fourier expansion. We have found that the eigenvalue distribution $\rho (\theta)$ is ungapped at low ($\mu, T$) and gapped at high ($\mu, T$), including the $\mu=0$ case (without the chemical potential term $S_{\textrm{g}}$). This suggests the existence of a phase transition in the fermionic case. We have also compared the critical points, and hence the phase diagram, between the bosonic and fermionic cases. 
The Monte Carlo simulation of the fermionic case entails a large CPU costs, due to the determinant (Pfaffian for $D=9$) after integrating out the fermionic degrees of freedom. Also, in the fermionic case the observables $\langle |u_n| \rangle$ are subject to a large finite-$\Lambda$ (cutoff parameter) effects at low temperature. These prevent us from making the similar analysis to the bosonic case numerically, and determining the nature of the phase transition. In the future it is instructive to study the nature of the phase transition of the fermionic case more closely. One strategy would be to study the eigenvalue distribution (\ref{eig_distribution}) at larger $N$ than $N=16$, which we have studied. Also, it is important to study the $D=9$ case, as well as the $D=3$ case, to see possible qualitative differences. We expect that at $D=9$, there is a phase transition between the gapped and ungapped phase, similarly to the $D=3$ case we have studied. This may have some connection to the Gregory-Laflamme instability of the black hole \cite{hep-th9301052}. To work on these interesting issues, we need to surmount the barrier of vast CPU costs. We hope to report on more analysis in future publications.

\acknowledgments
The authors would like to thank M. Hanada, G. Mandal, S. Minwalla, J. Nishimura and S.R. Wadia for valuable discussions and comments. T.A. and P.S. thank the warm hospitality at TIFR and ICTS respectively where part of this work was done. The work of T.A. was supported by MEXT Grant-in-Aid for Scientific Research (C) 17K05425, and "Priority Issue 9 on Post-K computerh (Elucidation of the Fundamental Laws and Evolution of the Universe). The numerical simulations were carried out at NTUA het clusters and FX10 at Kyushu University.

\appendix
\section{Review of the GWW phase transition} \label{GWW_appendix}
\begin{figure}[htbp]
%\vspace*{-15mm}
\centering % \begin{center}/\end{center} takes some additional vertical space
\includegraphics[width=7.4cm]{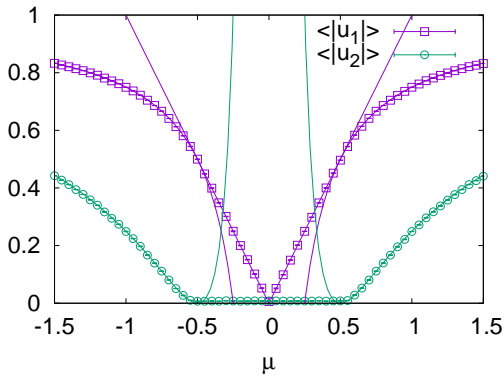}
\vspace{6mm}
\caption{\label{GWW_test} V.E.V's $\langle |u_{1,2}| \rangle$ for $N=128$ against $\mu$.}
%\vspace*{-15mm}
\end{figure}

In this section, we review the Gross-Witten-Wadia phase transition \cite{GWW1,GWW2,GWW3} of the unitary matrix model $S_{\textrm{g}}$, which is defined by eq. (\ref{chemical_pot}).  At large $N$, the V.E.V.'s of $|u_{1,2}|$, which is defined by eq. (\ref{u_n}), have been analytically calculated as 
\begin{eqnarray}
 \langle |u_1| \rangle = \left\{ \begin{array}{ll} \mu & \displaystyle{\left( 0 \leqq \mu \leqq \frac{1}{2} \right)} \\ \displaystyle{1 - \frac{1}{4\mu}} & \displaystyle{\left( \mu \geqq \frac{1}{2} \right)} \end{array} \right., \ \ 
% \langle |u_n| \rangle = \left\{ \begin{array}{ll} 0 & \displaystyle{\left( |\mu| \leqq \frac{1}{2} \right)} \\ \displaystyle{\left( 1 - \frac{1}{2\mu} \right) \left\{ \frac{P_n' \left( 1 - \frac{1}{\mu}\right)}{n(n+1)} + \frac{P_{n-1}' \left( 1 - \frac{1}{\mu}\right)}{n(n-1)}\right\} } & \displaystyle{\left( |\mu| \geqq \frac{1}{2} \right)} \end{array} \right. 
 \langle |u_2| \rangle = \left\{ \begin{array}{ll} 0 & \displaystyle{\left( 0 \leqq \mu \leqq \frac{1}{2} \right)} \\ \displaystyle{\left( 1 - \frac{1}{2 \mu} \right)^2 } & \displaystyle{\left( \mu \geqq \frac{1}{2} \right)} \end{array} \right. \label{GWW_un}
\end{eqnarray}
Hence we have
\begin{eqnarray}
 & &  \frac{d \langle |u_1| \rangle}{d \mu} = \left\{ \begin{array}{ll} 1 & \displaystyle{\left( 0 \leqq \mu \leqq \frac{1}{2} \right)} \\ \displaystyle{\frac{1}{4\mu^2}} & \displaystyle{\left( \mu \geqq \frac{1}{2} \right)} \end{array} \right., \ \  \frac{d^2 \langle |u_1| \rangle}{d \mu^2} = \left\{ \begin{array}{ll} 0 & \displaystyle{\left( 0 \leqq \mu \leqq \frac{1}{2} \right)} \\ \displaystyle{- \frac{1}{2 \mu^3} } & \displaystyle{\left( \mu \geqq \frac{1}{2} \right)} \end{array} \right. \label{GWW_un2}
\end{eqnarray}
At $\displaystyle \mu = \frac{1}{2}$, $\displaystyle \langle |u_1| \rangle$ and $\displaystyle  \frac{d \langle |u_1| \rangle}{d \mu}$ are continuous but $\displaystyle \frac{d^2 \langle |u_1| \rangle}{d \mu^2}$ (hence the third derivative of the free energy) is not continuous. This third-order phase transition is called the GWW phase transition.
%\begin{wrapfigure}{r}{70mm}
%\vspace*{-15mm}
%\centering % \begin{center}/\end{center} takes some additional vertical space
%\includegraphics[width=7.4cm]{wilson-0502227new2.eps}
%\vspace{4mm}
%\caption{\label{GWW_test} V.E.V's $\langle |u_{1,2}| \rangle$ for $N=128$ against $\mu$.}
%\vspace*{-15mm}
%\end{wrapfigure}
We take the static diagonal gauge (\ref{static_diagonal}). Adding the gauge-fixing term (\ref{gauge_fixing}), we apply the Metropolis algorithm to the action
\begin{eqnarray}
  S_{\textrm{g}} + S_{\textrm{g.f.}} = 2 N \mu \sum_{k=1}^{N} \cos \alpha_k - \sum_{k,l = 1, k \neq l}^{N} \log \sin \left| \frac{\alpha_k - \alpha_l}{2} \right|.  \label{GWW_action2}
\end{eqnarray}
We plot the VEVfs $\displaystyle \langle |u_{1,2}| \rangle$ against $\mu$ in fig. \ref{GWW_test} for $N=128$. Clearly, the result is invariant under flipping the sign as $\mu \to - \mu$, since this amounts to shifting $\alpha_k \to \alpha_k + \pi$ ($k=1,2,\cdots,N)$ all together, due to $\cos (x+\pi) = - \cos x$.

\end{document}